\documentclass[12pt]{article}

\usepackage[dvips]{graphicx}
\renewcommand{\baselinestretch}{1.75}
\font\scaps=cmcsc10 scaled\magstep1 % small capitals

\newcommand \be{\begin{equation}}
\newcommand \ee{\end{equation}}
\newcommand \ba{\begin{eqnarray}}
\newcommand \ea{\end{eqnarray}}
\begin{document}

\def\today{\ifcase\month\or
 January\or February\or March\or April\or May\or June\or
 July\or August\or September\or October\or November\or December\fi
 \space\number\day, \number\year}
%
%\line {Submitted \hfil \today}
%\hfil PostScript file created: \today

\hfil PostScript file created: \today{}; \ time \the\time \ minutes
\vskip .15in

%\setlength{\baselineskip}{22pt} %{8mm} 7pp.
%

% \centerline {INVERSE GAUSSIAN DISTRIBUTION }
% \centerline {AND ITS APPLICATION
% TO EARTHQUAKE OCCURRENCE
% }

\centerline {RANDOM STRESS AND OMORI'S LAW}

\vskip .15in
\begin{center}
{Yan Y. Kagan }
\end{center}
\centerline {Department of Earth and Space Sciences,
University of California,}
\centerline {Los Angeles, California 90095-1567, USA;}
\centerline {Emails: {\tt ykagan@ucla.edu,
kagan@moho.ess.ucla.edu }}
\vskip 0.02 truein

\vspace{0.15in}

\noindent
{\bf Abstract.}
We consider two statistical regularities that were used to
explain Omori's law of the aftershock rate decay: the L\'evy
and Inverse Gaussian (IGD) distributions.
These distributions are thought to describe stress behavior
influenced by various random factors: post-earthquake
stress time history is described by a Brownian motion.
Both distributions decay to zero for time intervals close to
zero.
But this feature contradicts the high immediate aftershock
level according to Omori's law.
We propose that these statistical distributions are influenced
by the power-law stress distribution near the earthquake focal
zone and we derive new distributions as a mixture of power-law
stress with the exponent $\psi$ and L\'evy as well as IGD
distributions.
Such new distributions describe the resulting inter-earthquake
time intervals and closely resemble Omori's law.
The new L\'evy distribution has a pure power-law form with the
exponent $-(1+\psi/2)$ and the mixed IGD has two exponents:
the same as L\'evy for small time intervals and $-(1+\psi)$
for longer times.
For even longer time intervals this power-law behavior should
be replaced by a uniform seismicity rate corresponding to the
long-term tectonic deformation.
We compute these background rates using our former analysis of
earthquake size distribution and its connection to plate
tectonics.
We analyze several earthquake catalogs to confirm and
illustrate our theoretical results.
Finally, we discuss how the parameters of random stress
dynamics can be determined through a more detailed statistical
analysis of earthquake occurrence or by new laboratory
experiments.

\vskip .15in
\noindent
{\bf Short running title}:
{\sc
% Inverse Gaussian distribution
Random stress and Omori's law
}


% \newpage
\vskip .25in

\section{Introduction}
\label{intro}

Our aim is to investigate the connection between random stress
and the Omori law of aftershock occurrence.
Several attempts to explain Omori's law have been published.
Since stress in the Earth's interiors cannot be easily measured
or calculated, these studies usually consider stress as a
scalar stochastic variable, ignoring for a while its tensorial
nature.
In this work we consider the stress as a scalar.

Here we use the scalar seismic moment $M$ directly, but for
easy comparison we convert it into an approximate moment
magnitude using the relationship
\be
m_W \ = \ {2 \over 3} \, (\log_{10}M - 9.0) \, ,
\label{Eq17a}
%\label{Gl_Eq01}
\ee
(Hanks, 1992), where $M$ is measured in Newton m (Nm).
Since we are using almost exclusively the moment magnitude,
later we omit the subscript in $m_W$.

Kagan (1982) considered stress time history as a Brownian
motion, i.e., randomly fluctuating stress acting on the
stressed environment of an earthquake focal zone.
Here the probability density for the time intervals
between earthquake events has a L\'evy distribution with a
power-law tail having the exponent $-3/2$.
Kagan and Knopoff (1987) added a tectonic drift component to
the Brownian motion.
For such a case, the distribution of inter-event times is
the Inverse Gaussian distribution (IGD) which, depending on
the value of the initial stress drop and velocity of tectonic
motion, can exhibit occurrence patterns ranging from the
L\'evy distribution to a quasi-periodic occurrence.

Matthews {\it et al.}\ (2002) also proposed the IGD as a law
for inter-earthquake intervals; however, they considered only
a limiting, quasi-periodic long-term distribution.
The authors suggest that
``... [the Inverse Gaussian] distribution has the following
noteworthy properties: (1) the probability of immediate
[large earthquake] rerupture is zero; (2) the hazard rate
increases steadily from zero at $t=0$ to a finite maximum near
the mean recurrence time ..." and real earthquake
occurrence follows the same pattern.
Ellsworth (1995), Ellsworth {\it et al.}\ (1999), Nadeau {\it
et al.}\ (1995), and Nadeau and Johnson (1998) present several
sequences of recurrent quasi-periodic earthquakes which, in
their opinion, confirm a smaller coefficient of variation than
the Poisson process for earthquake recurrence intervals.

There is ample evidence that after almost any earthquake
subsequent events follow Omori's law.
This has been observed for smaller earthquakes (aftershocks)
as well as large events comparable in size to the original
shock or even exceeding it (Kagan and Jackson, 1999).
Omori's law pattern of power-law rate decaying seismicity is
observed at decadal and centuries long scales (Utsu {\it
et al.}, 1995; Stein and Liu, 2009; Ebel, 2009).

Moreover, the presumed pattern of earthquake quasi-periodicity
depends on the argument that stress drops to zero or close
to zero in the focal zone of an earthquake.
Thus, it would be necessary to wait for some time before a
critical stress level is reached.
Kagan and Jackson (1999) presented several examples of large
earthquakes which followed after a short time interval (often
just a few days) within a focal zone of similar large events.
Such an inter-earthquake interval is clearly insufficient for
stress to be replenished by tectonic motion.
We present additional evidence for this pattern later in this
paper.

Thus, for relatively short time intervals, earthquakes are
clustered in time with the coefficient of variation
significantly greater than that for the Poisson process (Kagan
and Jackson, 1991).
Why is it widely believed that large earthquakes are
quasi-periodic in time?
Evidence mostly comes from specifically selected sequences
(Bakun and Lindh, 1985; Ellsworth, 1995) or from paleo-seismic
investigations (Ellsworth {\it et al.}, 1999).
However, paleo-seismic investigations have poor time
resolution: two or more events occurring closely in time are
likely to be identified as one event.
Hence, their coefficient of variation estimates should be
biased towards smaller values: a more periodic pattern.
Some paleo-seismic investigations (Marco {\it et al.}, 1996.
Rockwell {\it et al.}, 2000; Ken-Tor {\it et al.}, 2001;
Dawson {\it et al.}, 2003) also suggest that earthquakes
follow a long-term clustering pattern at least in regions of
slow tectonic deformation.

Additionally, paleo-seismicity studies the Earth's displacement
only at the surface.
But temporal and spatial distributions of earthquakes rupturing
the surface at one site substantially differ from general
earthquake statistics (Kagan, 2005).
The earthquake size distribution is also significantly
different for site- and area-based statistics.
Rigorously reliable statistical properties that are relevant
for theoretical studies as well as for seismic hazard
estimates can only be obtained by analyzing instrumental
earthquake catalogs.

Measurements from instrumental earthquake catalogs indicate
that short time intervals between large earthquakes are much
more frequent than even the Poisson model would suggest (Kagan
and Jackson, 1991; 1999).
Moreover, the problem of biased selection is also hard to
avoid in historical and paleo-seismic data: sequences not
displaying suggested patterns or with a small number of events
are less likely to be considered.

Recently, repeating microquakes have been studied in the
Parkfield area (Nadeau {\it et al.}, 1995; Nadeau and
Johnson, 1998, and references therein).
These event sequences exhibit certain properties of
characteristic quasi-periodic earthquakes: regularity of
occurrence and nearly identical waveforms.
They show many other interesting features of earthquake
occurrence.
However, these micro-events are not characteristic earthquakes
in a strict sense: they do not release almost all the tectonic
deformation on a fault segment while real characteristic
earthquakes are assumed to do so.

As with large characteristic earthquakes, attempts using
micro-earthquake quasi-periodicity for their prediction have
not yet succeeded.
Zechar, J. D. (private communication, 2010) attempted
to forecast repeating small earthquakes on the San Andreas
fault near Parkfield, based on event regularity.
While retrospective tests indicated that the forecasts based
on recurrence times were better than random, the forward tests
were unsuccessful.

Moreover, these micro-events apparently occur on isolated
asperities surrounded by creeping zones.
Therefore, such an asperity exists as a secluded entity and
may produce characteristic quasi-periodic rupture events,
similar to fracture in many laboratory experiments.
But tectonic earthquakes occur in an environment that is never
isolated.
This may explain their ubiquitous clustering in time and
space.

The quasi-periodic model for characteristic earthquakes (see
for example, McCann {\it et al.}, 1979; Bakun and Lindh, 1985;
Nishenko, 1991) has been put to a prospective test using new
data.
Testing the validity of this hypothesis ended in failure
(Rong {\it et al.}, 2003 and references therein; Bakun {\it et
al.}, 2005; Jackson and Kagan, 2006).

% 18:00-31-JAN-2010
% How can tectonic strain maps be used to investigate earthquake
How can we investigate earthquake occurrence patterns in
various tectonic zones?
Our statistical studies of seismicity (see for example, Kagan,
1991, 2007; Kagan and Jackson, 1991; Kagan {\it et al.}, 2010)
are biased because earthquake rate is dominated by places with
a high tectonic rate.
Thus, the likelihood function maximum is mainly determined by
these earthquakes.
However, we should study earthquake patterns in continental
areas (active and non-active) where the earthquake rate is
low, but the vulnerable human population is large (see Stein
and Liu, 2009; Parsons, 2009).
A naive extrapolation of aftershock sequence rates would
exaggerate seismic hazard.
Instead, we need a convincing tool to produce a truer
estimate.

Our recent results suggest that earthquake occurrence can be
modeled by the Poisson cluster process: the combination of the
long-term rate caused by tectonic strain with short-term
clustering.
For large and medium strain rates, the model appears to work
well (see again Kagan {\it et al.}, 2010).
The model is producing reasonable long- and short-term
forecasts of earthquake occurrence both for local
(Helmstetter {\it et al.}, 2006) and global (Kagan and
Jackson, 2010) seismicity.
But to extrapolate such a model to small tectonic rates we
would need additional arguments; we must extract the rates by
comparing seismicity patterns with tectonic strain maps.
% However, it is not clear how to obtain this information.
Bird {\it et al.}\ (2010) discuss the dependence of the
long-term inter-earthquake rate on strain.
We need to show how the short-term properties of earthquakes
behave in the regions of varying strain.
As shown below, the transition time from Omori's law decay to
a quasi-stationary rate clearly increases with a decrease in
the strain rate.
% Below, we obtain quantitative description of earthquake
% occurrence parameters in diverse tectonic zones.
As the result of these studies, we propose that the same
statistical model can be applied for regions of both high and
low tectonic strain.

\section{Aftershock temporal distribution, theoretical
analysis }
\label{analys}

\subsection{L\'evy distribution }
\label{analys1}

Kagan (1982) proposed a heuristical model for an earthquake
fracture as follows.
At the moment that any given earthquake rupture
stops, the stress near the edge of the rupture is lower than
the critical breaking stress for extension of the fracture.
The subsequent stress history near the earthquake rupture tip
depends on other fractures in the neighborhood and additional
random factors.
In this case, the time history of the stress might resemble a
Brownian random walk.
The stress-time function is thus given as a solution to the
diffusion equation.
When the stress reaches a critical threshold level, a new
earthquake begins.
The distribution density of the time intervals of the `first
passage' (Feller, 1966) is the L\'evy distribution
\be
f(t) = { { \sigma } \over \sqrt{ 2 \, \pi D \, t^3 } } \,
\exp \left (- \, {{\sigma^2} \over {2 D \, t}} \right )
\, ,
\label{Eq01}
\ee
where $D$ is the diffusion coefficient, $t$ is the
time, and $\sigma$ is the threshold or barrier stress:
the difference between initial breaking stress and that
when rupture ceases.
This distribution as a function of stress is the Rayleigh law;
as a function of time it is the L\'evy distribution ({\sl cf}.
Zolotarev, 1986).
In this model, the stress is taken to be a scalar, which
corresponds to the addition of perfectly aligned stress
tensors; the interaction of misaligned tensors will be
discussed in a later work.
The L\'evy distribution (1) was used by Kagan (1982,
see Eqs.~8 and 9) to model the time dependent part of the
space-time sequence of micro-earthquakes.
Here we are address only the time sequence of events.

\begin{figure}
\begin{center}
\includegraphics[width=0.75\textwidth]{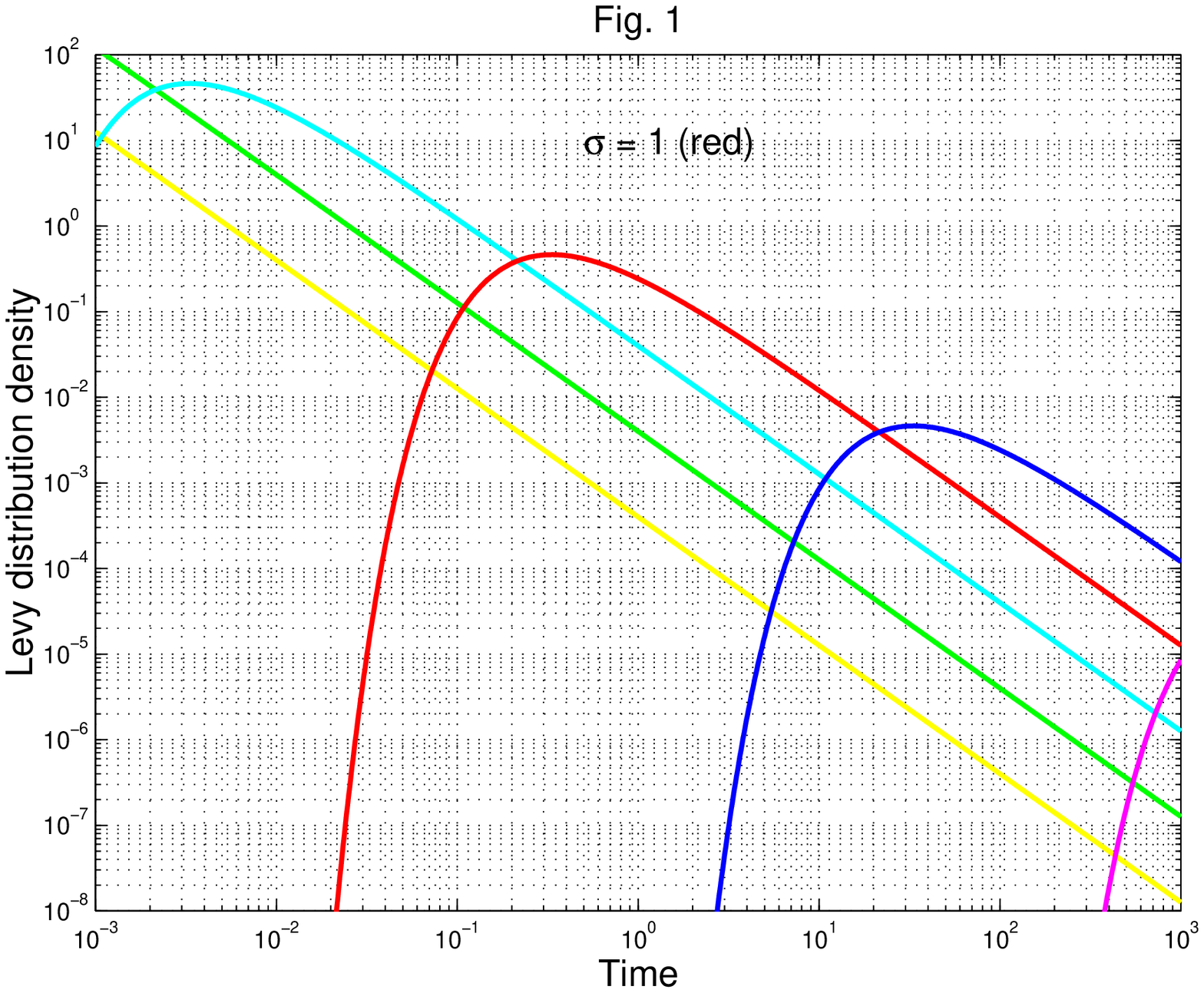}
% #fig01 IGD3.M P52,p.72
\caption{\label{fig01}
}
\end{center}
Plot of PDFs for the L\'evy distribution (\ref{Eq01}),
$D = 1.0; \ \sigma = 0.001, 0.01, 0.1, 1.0, 10.0, 100.0$.
\end{figure}

Fig.~\ref{fig01} displays several L\'evy curves for various
values of stress drop, $\sigma$.
The curve's tail is a power-law similar to that exhibited by
Omori's law.
However, for small values of time the curves decay to zero:
depending on the stress drop, Brownian motion takes time to
reach a critical stress level.
This latter feature, though observed in aftershock sequences,
is most likely caused by effects from coda waves of both the
mainshock and stronger aftershocks that prevent identifying
smaller aftershocks (Kagan, 2004).
This opinion is supported by aftershock registration in the
high-frequency domain (Enescu {\it et al.}, 2009) where their
time delay is substantially reduced.
Measurements of total seismic moment rate decay in aftershock
sequences (Kagan and Houston, 2005) suggest that aftershocks
start right after the end (within one minute for a $M6.7$
earthquake) of a mainshock rupture.
Therefore, the left-hand decay of the curves in
Fig.~\ref{fig01} is not realistic.

There is another problem in comparison of theoretical
distribution with aftershock observations.
The L\'evy curves may explain the inter-event time statistics
only for the first generation (parent-child) of clustered
earthquakes.
Most observed aftershocks or clustered events are
generally separated from their `parents' by multiple
generations.
For example, the present aftershocks of the 1811-12 New Madrid
earthquakes or the 1952 Kern County, California, earthquake
are mostly distant relatives of their progenitors
(mainshocks).
% The time distribution behaves differently: this can be
% seen from the exponent values of Omori's law (about 1.0) and
% the L\'evy distribution (1.5).
In the Omori law we count all aftershocks, disregarding their
parentage, whereas the L\'evy law describes the time
distribution for the {\sl next} event: when the stress level
reaches a critical value to trigger a new rupture.
When we count all events, we decrease the exponent from 1.5 to
1.0 (Kagan and Knopoff, 1987).

We assume that the next earthquake may occur in any place
in the focal zone of a previous event.
Attempts to localize the initiation of strong earthquakes have
not been successful; for example, the 2004 Parkfield
earthquake occurred well outside the previously specified area
(Jackson and Kagan, 2006).

We also assume that the stress drop after an earthquake has a
power-law distribution: it is proportional to
$\sigma^{-1-\psi}$, for $ 0 \le \psi < 1.0$.
Kagan (1994), Marsan (2005), Helmstetter {\it et al.}\ (2005),
Lavall\'ee (2008), Powers and Jordan (2010) support this
statement.

Then for the L\'evy distribution we obtain
\be
\phi \, (t) \, \propto \,
\int\limits_0^\infty f(t) \, \sigma^{-1-\psi} d \sigma
\, \, = \, \,
{ 1
\over 2 \, t \, \sqrt \pi \, (2 \, t \, D)^{\psi/2} }
\, \Gamma \left ( {{1-\psi} \over 2} \right )
\, ,
\label{Eq02}
\ee
where $\Gamma $ is a gamma function.
The new modified L\'evy distribution density function is a
power-law with the exponent $-(1+\psi/2)$.
The density should be truncated at the left side, because
aftershocks cannot be observed at short-time intervals close
to a mainshock (Kagan, 2004).
Normalizing the distribution would depend on this truncation.
Kagan (1991, Eq.~8) introduced a minimum time ($t_M$), where
$M$ is a seismic moment of a preceding event (see also
Eq.~\ref{Eq22} below).
Before that time, the aftershock rate is measured with a
substantial undercount bias (Kagan, 2004).

\subsection{Inverse Gaussian distribution
}
\label{analys2}

As Matthews {\it et al.}\ (2002) indicate, the Inverse Gaussian
Distribution (IGD) has been known since 1915.
However, this distribution only acquired its name in 1945.
This is the name found in recent statistical encyclopedias and
books (for example, Kotz {\it et al.}, 2006; Seshadri, 1999;
Chhikara and Folks, 1989), on the Wolfram website, and in
Wikipedia, while the ``Brownian passage-time" (BPT) is little
known.
Kagan and Knopoff (1987) proposed to use this distribution
(without calling it IGD) to describe the inter-earthquake time
distribution.
Their major impetus was to explain aftershock statistics
(Omori's law) by stress dynamics.

\begin{figure}
\begin{center}
\includegraphics[width=0.75\textwidth]{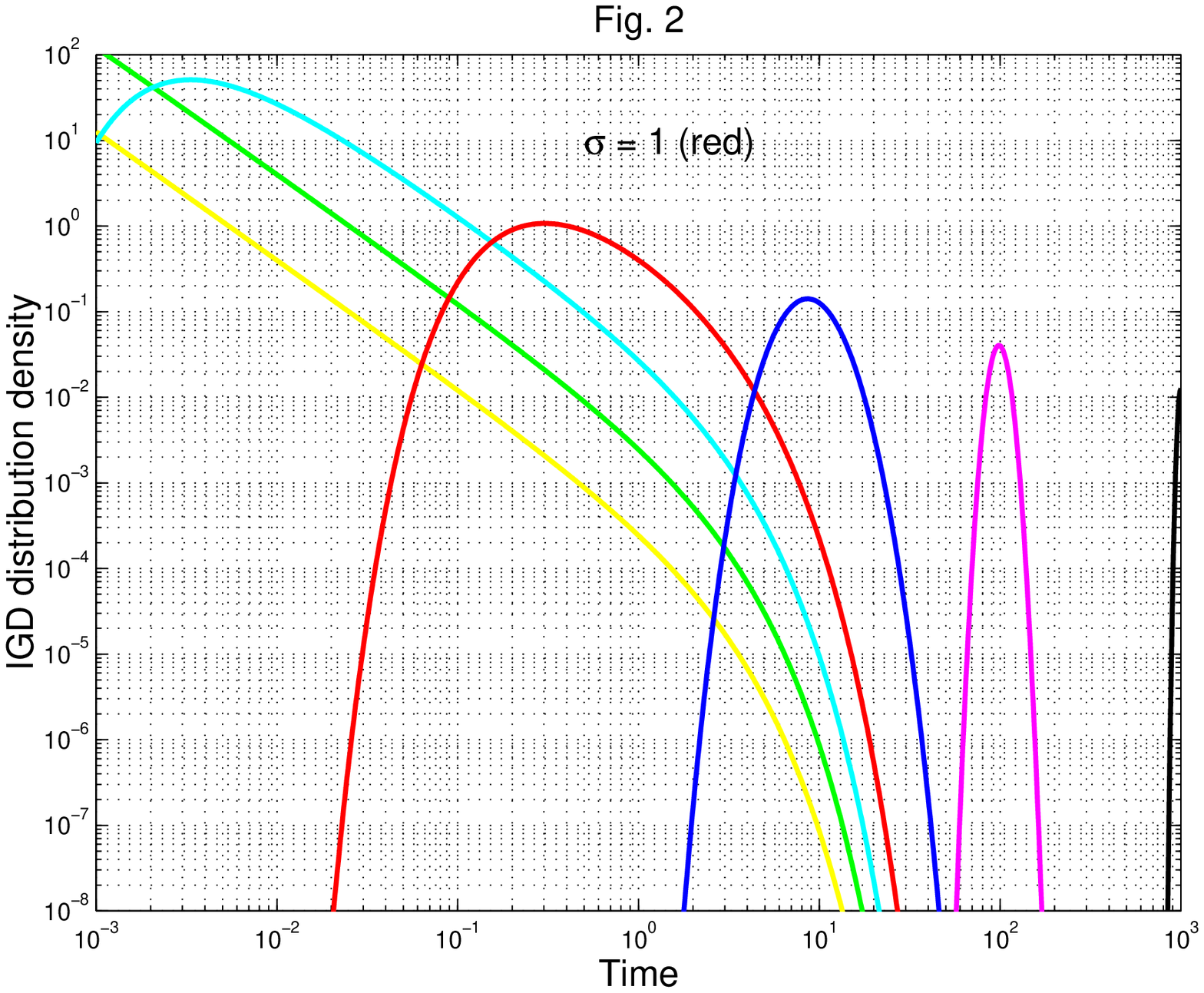}
% #fig02 IGD3.M P52,p.72
\caption{\label{fig02}
}
\end{center}
Plot of PDFs for the IGD distribution (\ref{Eq03}),
$V = \sqrt D; \ D = V^2; \ \sigma = 0.001, 0.01, 0.1, 1.0,
10.0, 100.0, 1000.0$.
\end{figure}

Here we discuss the appropriate pairwise interval law
for a model in which a steadily increasing source of stress,
which we take to be due to plate tectonics, is added to or
subtracted from a random or diffusion component, where the
distribution (\ref{Eq01}) describes the density of earthquake
recurrence times in the absence of tectonic loading.
% Under the model of Brownian motions in the presence of an
% external steady field, the frequency of occurrence law will be
% asymptotic to the Omori law for short time intervals.
% There is no guarantee that the familiar exponential law for
% the time intervals between Poisson random events will be the
% result.
If the rate of tectonic loading is a constant {\sl V}, the
distribution density $f(t)$ is modified to (Feller, 1966,
p.~368) become the Inverse Gaussian distribution
\be
f(t) = { { \sigma } \over \sqrt{ 2 \, \pi D \, t^3 } } \, \exp
\left [- \, {{(\sigma-V \, t)^2} \over {2 D \, t}} \right ]
\, .
\label{Eq03}
\ee
For tectonic loading velocity $V = 0$, this equation
transforms to (\ref{Eq01}).
Fig.~\ref{fig02} displays several IGD curves for various
values of stress drop.
For small values of $\sigma$ and $t$, the curves are similar
to the values of L\'evy distribution (\ref{Eq01}): if the
stress drop is insignificant, or the tectonic loading
influence is small in initial stages of stress behavior.

Assuming again that the stress in a fault neighborhood is
distributed according to a power-law, we obtain a new modified
distribution of inter-earthquake time, based on the IGD
(\ref{Eq03})
\ba
\phi \,(t) & \propto &
\int\limits_0^\infty f(t) \, \sigma^{-1-\psi} d \sigma
% \nonumber\\
% \, & = & \,
\ = \
{ {\exp \left [ { - V^2 \, t / (2 D) } \right ] }
\over 2 \, t \, \sqrt \pi \, (2 \, t \, D)^{\psi/2} }
\ \Biggl [
\, \Gamma \left ( {{1-\psi} \over 2} \right )
\, {_{1}F_{1}} \,
\left ( {{ 1-\psi} \over 2}, \, {1 \over 2}, \, { { V^2 \, t
} \over {2 D} } \right )
\nonumber\\
& + &
V \, \sqrt { { { 2 \, t } \over D} }
\, \, \Gamma \left ( 1- {\psi \over 2} \right )
\, {_{1}F_{1}} \,
\, \left ( 1-{\psi \over 2}, \, {3 \over 2}, \, { { V^2 \, t
} \over {2 D} } \, \right )
\Biggr ]
\, ,
\label{Eq04}
\ea
where $ \, {_{1}F_{1}} \, $ is a Kummer confluent
hypergeometric function (Wolfram, 1999; Abramowitz and Stegun,
1972, p.~504).
Another expression for the density is
\be
\phi \,(t) \ \propto \
{ {\exp \left [ { - V^2 \, t / (2 D) } \right ] }
\over 2 \, t \, \sqrt \pi \, (2 \, t \, D)^{\psi/2} }
\, \, \, \Gamma \left ( {{1-\psi} \over 2} \right )
\, \Gamma \left ( 1- {\psi \over 2} \right )
{ \, U }
\left ( {{ 1-\psi} \over 2}, \, {1 \over 2}, \, { { V^2 \, t
} \over {2 D} } \right )
\, ,
\label{Eq05}
\ee
where $\, U \, $ a confluent hypergeometric function ({\it
ibid.}).
As in Eq.~\ref{Eq02}, both of the above distributions should
be truncated as $t \to 0$ and can be normalized after it.

For certain values of $\psi$ the expressions (\ref{Eq04} and
\ref{Eq05}) can be simplified.
For example for $\psi = 0$
\be
\phi \, (t) \, \propto \,
{1 \over {2 t}} \left [ \, 1 + {\rm erf} \left ( V \sqrt { { t
\over {2 D} } } \, \right ) \right ]
\, .
\label{Eq06}
\ee
For $\psi = 0.5$ we obtain two equations.
For a positive $V$, using Eq.~13.6.3 by
Abramowitz and Stegun (1972), we transform (\ref{Eq04}) into
\be
\phi \, (t) \, \propto \,
{1 \over {2 \, t}} \, \, \sqrt { {\pi \, V} \over {4 D}} \,
\, \exp \left ( - { {V^2 \, t} \over { 4 D} } \right )
\left [ I_{-1/4} \left ( { {V^2 \, t} \over { 4 D} } \right ) +
I_{1/4} \left ( { {V^2 \, t} \over { 4 D} } \right ) \right ]
\, ,
\label{Eq07}
\ee
where $ I_{-1/4} $ and $I_{1/4} $ are modified Bessel
functions (Wolfram, 1999; Abramowitz and Stegun, 1972,
p.~374).

For a negative $V$ using Eq.~13.6.21 by Abramowitz and Stegun
(1972), we transform (\ref{Eq05}) into
\be
\phi \, (t) \, \propto \,
{1 \over {2 t} } \, \, \sqrt { - \, { {V} \over {\pi D}} } \,
\, \exp \left ( - \, { {V^2 \, t} \over { 4 D} } \right )
K_{1/4} \left ( { {V^2 \, t} \over { 4 D} } \right )
\, ,
\label{Eq08}
\ee
where $K_{1/4} $ is a modified Bessel function (Wolfram, 1999;
Abramowitz and Stegun, 1972).

\begin{figure}
\begin{center}
\includegraphics[width=0.75\textwidth]{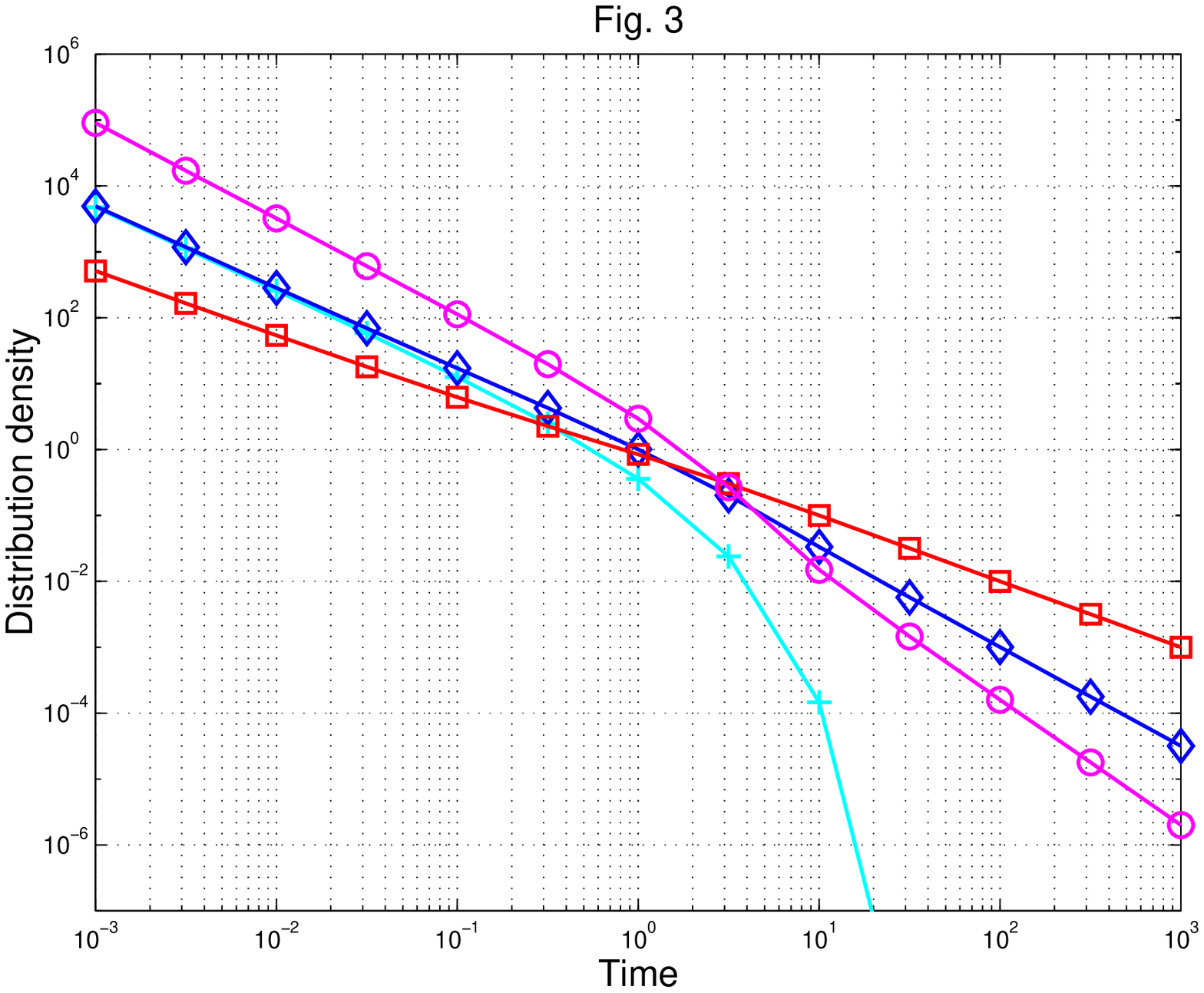}
% #fig03 IGD4.M P52,p.77
\caption{\label{fig03}
}
\end{center}
Plot of PDFs for the IGD distribution (Eqs.~\ref{Eq04} --
\ref{Eq08}).
% , $D = V^2$.
\hfil\break
Curve with squares (\ref{Eq06}): $V = \sqrt D, \ \psi = 0.0$;
\hfil\break
curve with diamonds (\ref{Eq07}): $V = \sqrt D, \ \psi = 0.5$;
\hfil\break
curve with crosses (\ref{Eq08}): $V = - \sqrt D, \ \psi =
0.5$;
\hfil\break
curve with circles (\ref{Eq04}): $V = \sqrt D, \ \psi = 0.9$.
\end{figure}

In Fig.~\ref{fig03} we display the new IGD curves for various
values of the $\psi$ parameter.
Although the general behavior of the curves remains power-law,
the curves change their slope at the
% scaled
time value of
about 1.0.
The curves with $V = \pm \sqrt D$ show the distribution
difference for tectonic loading sign; in the positive case
tectonic movement is opposite to the fault displacement during
an earthquake, whereas the negative sign corresponds to
motion consistent with the earthquake mechanism.
In the latter case, random fluctuations can bring the fault
to rupture only during the early period of development.

\begin{figure}
\begin{center}
\includegraphics[width=0.75\textwidth]{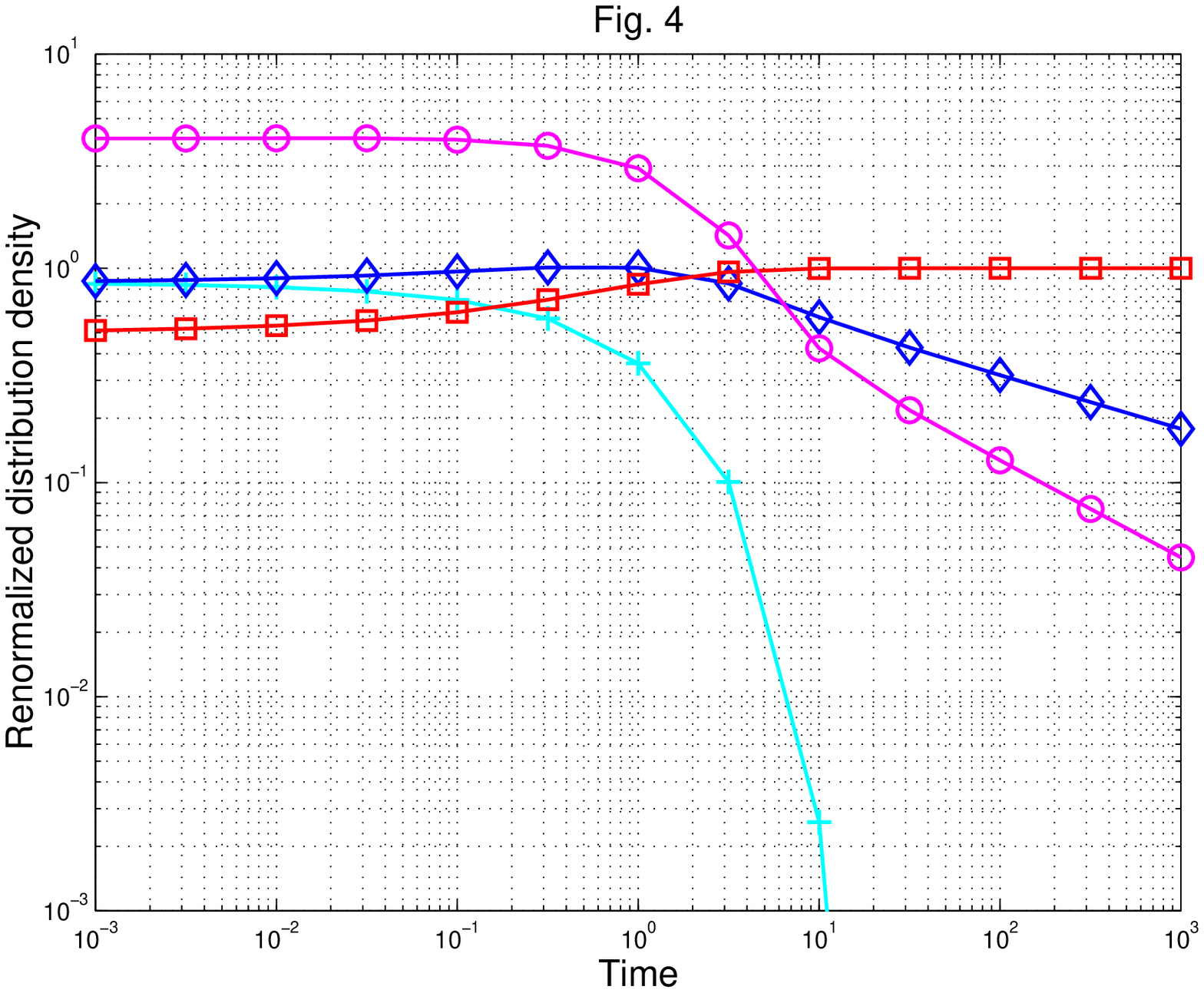}
% #fig04 IGD4.M P52,p.77
\caption{\label{fig04}
}
\end{center}
Plot of PDFs for the IGD distribution (Eqs.~\ref{Eq04} --
\ref{Eq08}), multiplied by $t^{1+\psi/2}$, $D = 1.0$.
\hfil\break
Curve with squares (\ref{Eq06}): $V = \sqrt D, \ \psi = 0.0$;
\hfil\break
curve with diamonds (\ref{Eq07}): $V = \sqrt D, \ \psi = 0.5$;
\hfil\break
curve with crosses (\ref{Eq08}): $V = - \sqrt D, \ \psi =
0.5$;
\hfil\break
curve with circles (\ref{Eq04}): $V = \sqrt D, \ \psi = 0.9$.
\end{figure}

To show the curves' behavior more clearly in Fig.~\ref{fig04},
the PDF values are multiplied by $t^{1+\psi/2}$.
For small time intervals curves are horizontal, suggesting
that the modified IGD is similar to the L\'evy distribution in
Eq.~\ref{Eq02}.

For small time values, the power-law exponents in
Fig.~\ref{fig03} are essentially the same as for compounded
L\'evy law (\ref{Eq02}).
This can be seen from modified Bessel function approximations
for small values of the argument (Abramowitz and Stegun, 1972,
Eqs.~9.6.7 and 9.6.9),
\be
I_{-1/4} (t) \, \propto \, \left ( {2 \over t} \right )^{1/4}
\ \ {\rm and} \ \
K_{1/4} (t) \, \propto \, \left ( {2 \over t} \right )^{1/4}
\, .
\label{Eq09}
\ee
In the general case the same result can be obtained with
Eq.~13.5.5 by Abramowitz and Stegun (1972).
Then for $t \to 0$ Eq.~\ref{Eq04} transforms into
\be
\phi \, (t) \, \propto \, t^{ \, - \, 1 \, - \, \psi/2}
\, .
\label{Eq10}
\ee

It is obvious from Figs.~\ref{fig03} and \ref{fig04} that,
except for the $\psi = 0$ curve, the slope of curves for
large values of the
% scaled
time increases when compared
with $t$ close to zero.
Using Eq.~13.1.27 of Abramowitz and Stegun (1972) we add
an exponential term in (\ref{Eq04}) within the hypergeometric
function $ \, {_{1}F_{1}} \, $.
Then for $t \to \infty$ we obtain
\be
\phi \, (t) \, \propto \, t^{ \, - \, 1 \, - \, \psi}
\, ,
\label{Eq11}
\ee
(Abramowitz and Stegun, 1972, Eq.~13.5.1).
Therefore, in Eq.~\ref{Eq07} the exponent would be $-1.5$
(Abramowitz and Stegun, 1972, Eq.~9.7.1),
whereas in Eq.~\ref{Eq08} the power-law term (Abramowitz and
Stegun, 1972, Eq.~9.7.2) is multiplied by an exponential decay
term
\be
\phi \, (t) \, \propto \, t^{ \, - 1.5}
\, \exp \left ( - \, { {V^2 \, t} \over { 4 D} } \right )
\, .
\label{Eq12}
\ee
See Figs.~\ref{fig03} and \ref{fig04}.

\section{Temporal distribution of aftershocks: Observations
}
\label{observ}

Many observations of Omori's law behavior have been published
(see Utsu {\it et al.}, 1995 and references therein).
There are some problems with these measurements.
The standard interpretation of the Omori law is that all
aftershocks are caused by a single mainshock.
However, the mainshock is often followed by strong aftershocks
and those are clearly accompanied by their own sequence of
events, and so on.
The second problem is that some earthquakes have very few or
no aftershocks; such sequences cannot be included in the
naive study of earthquake sequences, but can be analyzed as
stochastic processes.

Thus, three techniques can be applied to study the temporal
distribution of real earthquakes: 1) traditional,
phenomenological techniques based on observing individual
aftershock sequences; 2) using statistical moments of
earthquake occurrence, considered as a point process; 3)
applying stochastic process modeling to infer the parameter
values of earthquake temporal interaction.

\subsection{Aftershock sequences
}
\label{observ0}

Beginning with Omori (1894), the temporal distribution of
aftershock numbers has been studied for more than one hundred
years (Utsu {\it et al.}, 1995).
Aftershock rate decay is approximated as $t^{-p}$ with
parameter $p$ value close to 1.0 (Kagan, 2004).

But the simple, superficial study of aftershock rate decay
often encounters serious problems.
First, only relatively rich aftershock sequences can be
investigated by direct measurements; if there are too few
aftershocks, their properties can be studied only by combining
(stacking) many sequences.
Second, to isolate individual sequences one should exclude any
cases when one sequence is influenced by another, an
arbitrary procedure which may introduce a selection bias.
Third, an aftershock sequence often contains one or several
large events which are clearly accompanied by a secondary
aftershock sequence.
Taking the influence of secondary earthquakes into account is
not simple (see Section~\ref{observ1} for more detail).
Fourth, some sequences start with a strong foreshock which
is sometimes only slightly weaker than a mainshock.
Again, handling this occurrence presents serious problem.

Therefore, strong bias may result from directly measuring
Omori's law exponents.
Two other statistical methods considered below enable analysis
of the whole earthquake occurrence as a point process, to
minimize the problem of data and interpretation technique
selection bias.

\subsection{Temporal distribution for earthquake pairs }
\label{pair}

Kagan and Jackson (1991) investigated space-time pairwise
earthquake correlation patterns in several earthquake catalogs.
They showed that earthquake pairs follow a power-law
distribution for small time and distance intervals.
Kagan and Jackson (1999) showed that contrary to the seismic
gap model the clustering pattern continues for strong ($m \ge
7.5$) earthquakes.
Large shallow earthquakes can re-occur after a small time
interval.

Table~\ref{Table2} shows the location and focal mechanism
difference for $m \ge 7.5$ global shallow earthquakes in the
GCMT catalog (Ekstr\"om {\it et al.}, 2005; Ekstr\"om, 2007)
from 1976-2010.
The table format is similar to Table~1 in Kagan and Jackson
(1999).
However, here we kept only those pairs in the table
for which their focal zone overlap (the $\eta$-parameter) is
greater than 1.0:
\be
\eta = {{L_1 + L_2} \over {2 R}},
\label{Eq19}
\ee
where $L_1$ and $L_2$ are the respective rupture lengths for
the first and second earthquakes in the pair (see Eq.~3 in
Kagan and Jackson, 1999) and $R$ is the
distance between the centroids.
Therefore, if $\eta \ge 1.0$ the earthquake focal zones
would intersect.
For several doublets $\eta \ge 2$, implying that the smaller
event should be largely within the focal zone of the larger
earthquake.
Inspecting the time difference and the 3-D rotation angle
between focal mechanisms suggests that these high $\eta$ pairs
may occur after very short time intervals and have very
similar double-couple mechanisms.

All earthquakes in the table occur in subduction zones as
defined in Kagan {\it et al.}\ (2010).
However, even with relatively high deformation velocity at
these plate boundaries, the inter-earthquake time is in most
cases substantially lower than the time necessary for tectonic
motion to restore the critical stress conditions by the
occurrence time of the second earthquake (see the last column
in Table~1 by Kagan and Jackson, 1999).

\begin{figure}
\begin{center}
\includegraphics[width=0.75\textwidth]{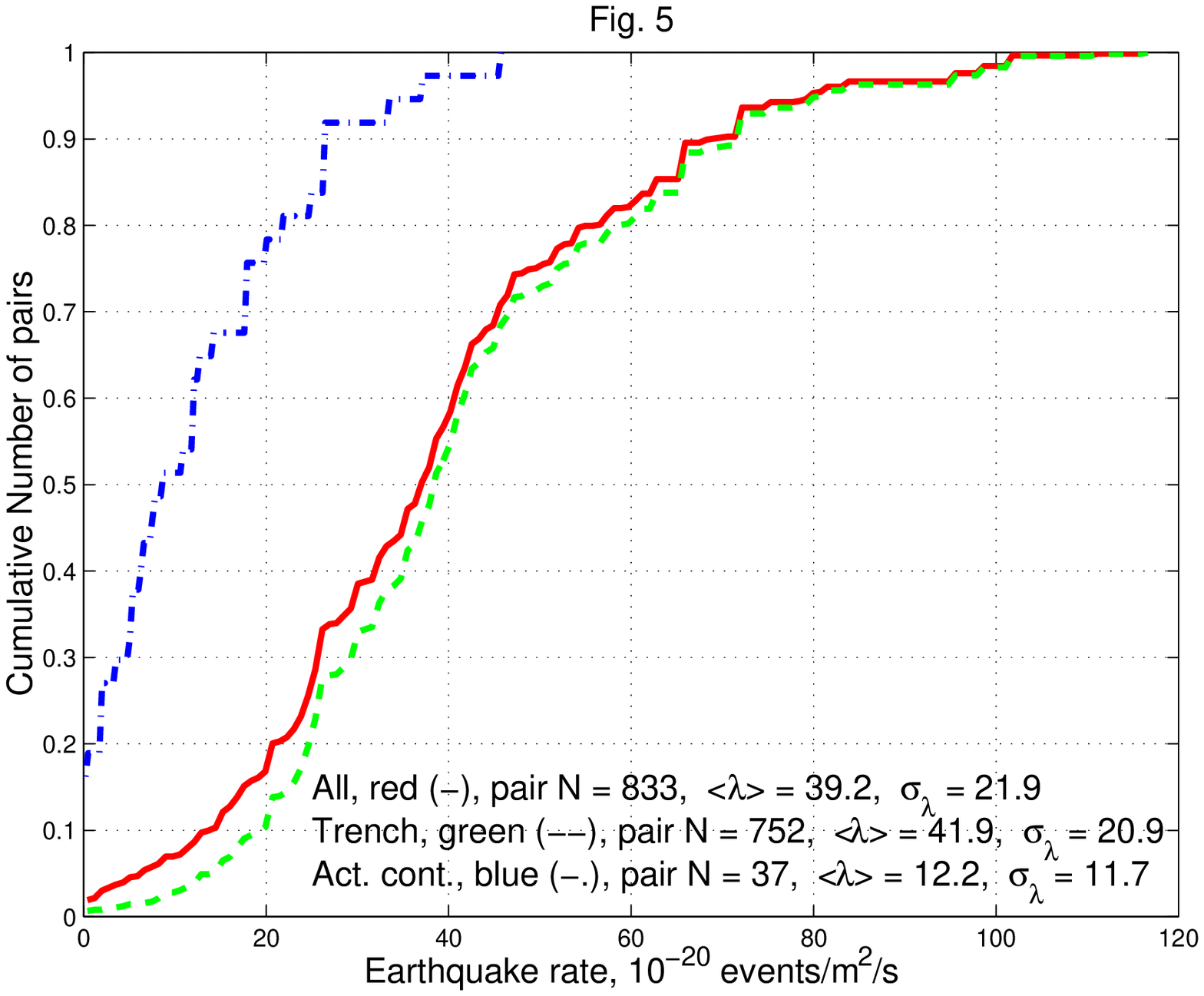}
% #fig07 TECTP3.M P53,p.60a; P54,p.11
\caption{\label{fig07}
}
\end{center}
Normalized cumulative distribution of earthquake rate for $m
\ge 6.5$ shallow earthquake pairs in all zones, trench
(subduction) zones, and active continental zones.
The pair numbers $N$, average rate $< \lambda >$ and its
standard deviation $\sigma_\lambda$ are also shown.
Earthquake rate is taken from a table by Bird {\it et al.}\
(2010) for a magnitude threshold $m_t = 5.66$.
\end{figure}

Fig.~\ref{fig07} shows how the normalized number of $m \ge
6.5$ shallow earthquake pairs depends on the tectonic
deformation rate as defined by Bird {\it et al.}\ (2010).
Three curves are shown: all earthquakes from the GCMT catalog,
earthquakes from subduction zones (Kagan {\it et al.}, 2010),
and events from active continental zones.

\begin{figure}
\begin{center}
\includegraphics[width=0.75\textwidth]{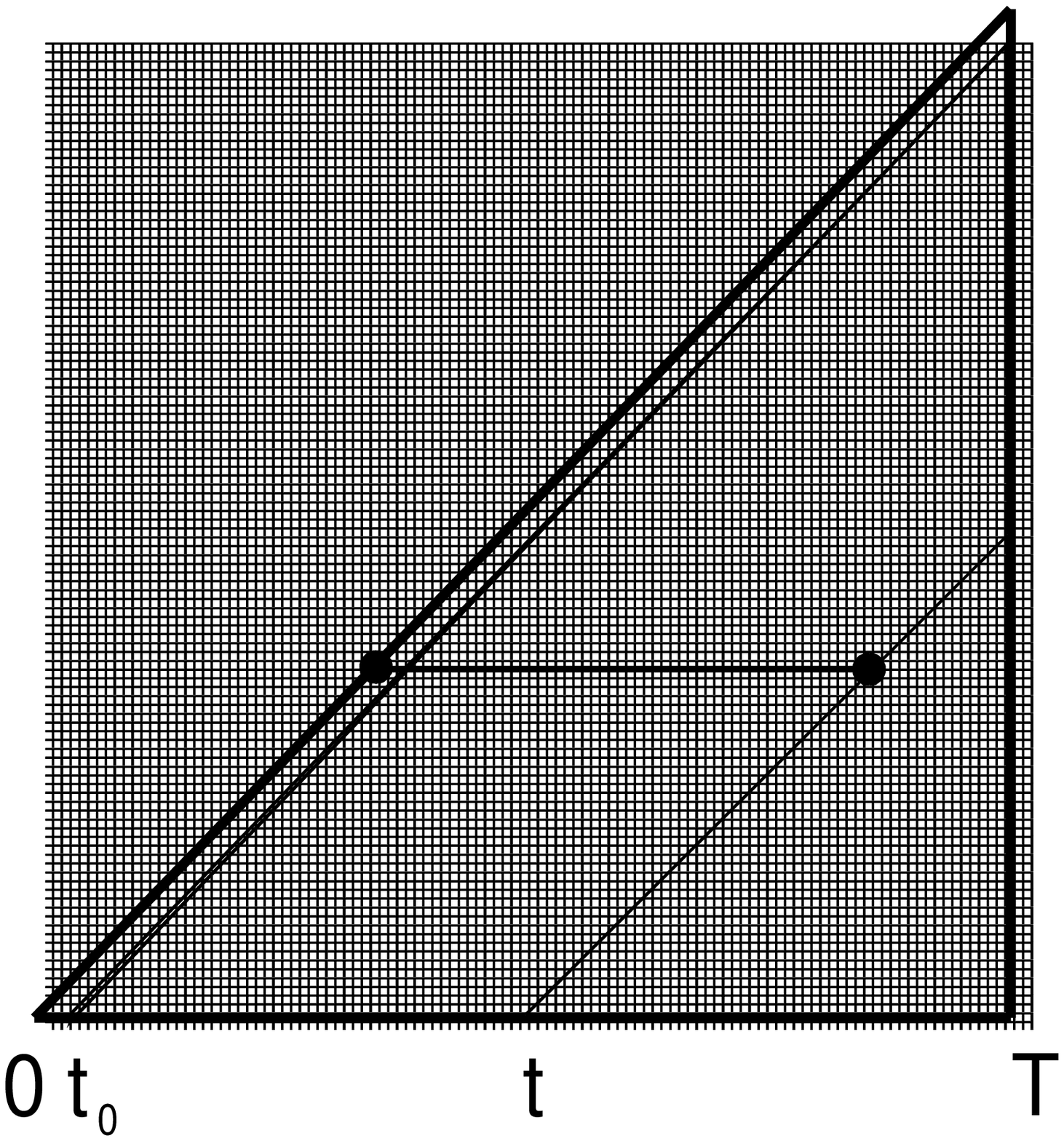}
% #fig07a IGD2.UIS P53,p.61
\caption{\label{fig07a}
}
\end{center}
Integration domain for calculating inter-earthquake rates
in a catalog of duration $T$.
The minimum time interval, $t_0$, corresponds to the coda
duration time of the first event.
\end{figure}

Fig.~\ref{fig07a} shows the integration domain for calculating
earthquake pair rates.
The diagram is a square with a side length equal to a catalog
duration, $T$; since the plot is symmetric, only the
lower-right portion of the square is shown.
The first event shown as a filled circle, is supposed to be at
the square diagonal, the second one at the end of a hatched
area.
We assume that the time difference between earthquakes cannot
be less than $t_0$ (similar to $t_M$ in Eq.~\ref{Eq22}).

For the Poisson process, the interval pair density is uniform
(see Kagan and Jackson, 1991, Eq.~1).
Thus, the rate is proportional to the hatched area.
For the normalized survival rate
\be
n_p \ = \ { \left ( { T - t } \over {T - t_0} \right)^2 } \, ,
\label{Eq20}
\ee
where $t_0$ is the minimum time interval, $T$ is the catalog
duration and $t$ is the inter-earthquake time interval.

For the power-law time distribution with distribution density
$ \phi (t) \ \propto \ t^{- 1 - \theta} $,
we obtain the normalized survival rate by integrating over the
domain shown in Fig.~\ref{fig07a}:
\be
n_p \ = \ { \left ( { t^{-\theta} - T^{-\theta} } \over
{ t_0^{-\theta} - T^{-\theta} } \right) } \, .
\label{Eq21}
\ee

\begin{figure}
\begin{center}
\includegraphics[width=0.75\textwidth]{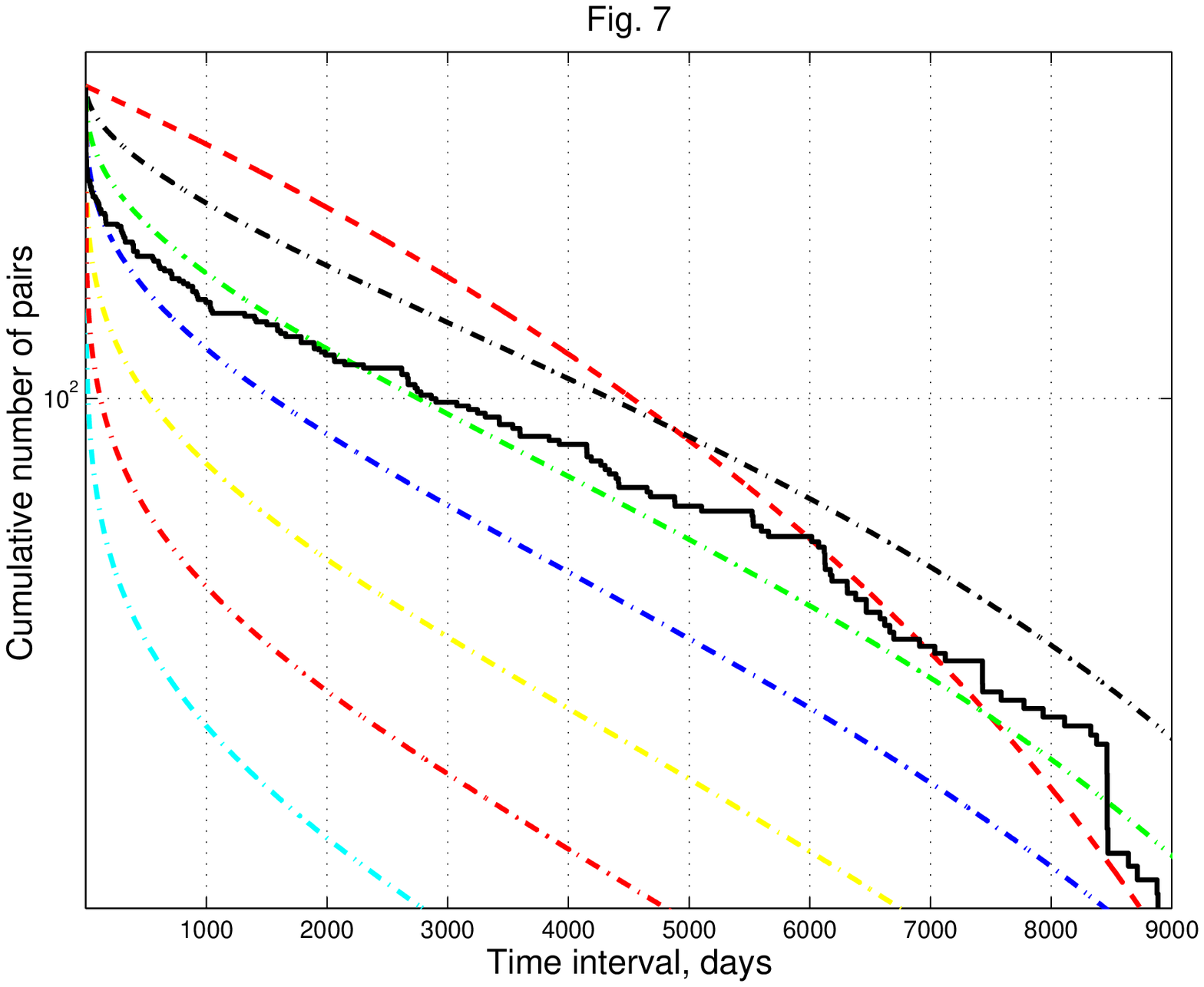}
% #fig08 FPSM2L.FOR BIRD_TECT2.FOR TIMED1_650_TF.M P53,p.61
\caption{\label{fig08}
}
\end{center}
Earthquake pair numbers in all zones with rates
$0-26 \times 10^{-20}$~events/m$^2$/s.
Earthquake rate is taken from a table by Bird {\it et al.}\
(2010) for a magnitude threshold $m_t = 5.66$.
Pair number is 240.
Solid curve is earthquake pair numbers;
dashed curve is Poisson approximation (\ref{Eq20}),
dash-dotted curves are for power-law approximations
(\ref{Eq21}),
the $\theta$-value is 0.5, 0.65, 0.75, 0.85, 0.925, and 0.99
from top to bottom.
The catalogs time interval is 1976/1/1--2010/11/14.
Average recurrence time ($\bar t$) for earthquakes is $\bar t
\ = \ 3272 \pm 3596$ days.
\end{figure}

\begin{figure}
\begin{center}
\includegraphics[width=0.75\textwidth]{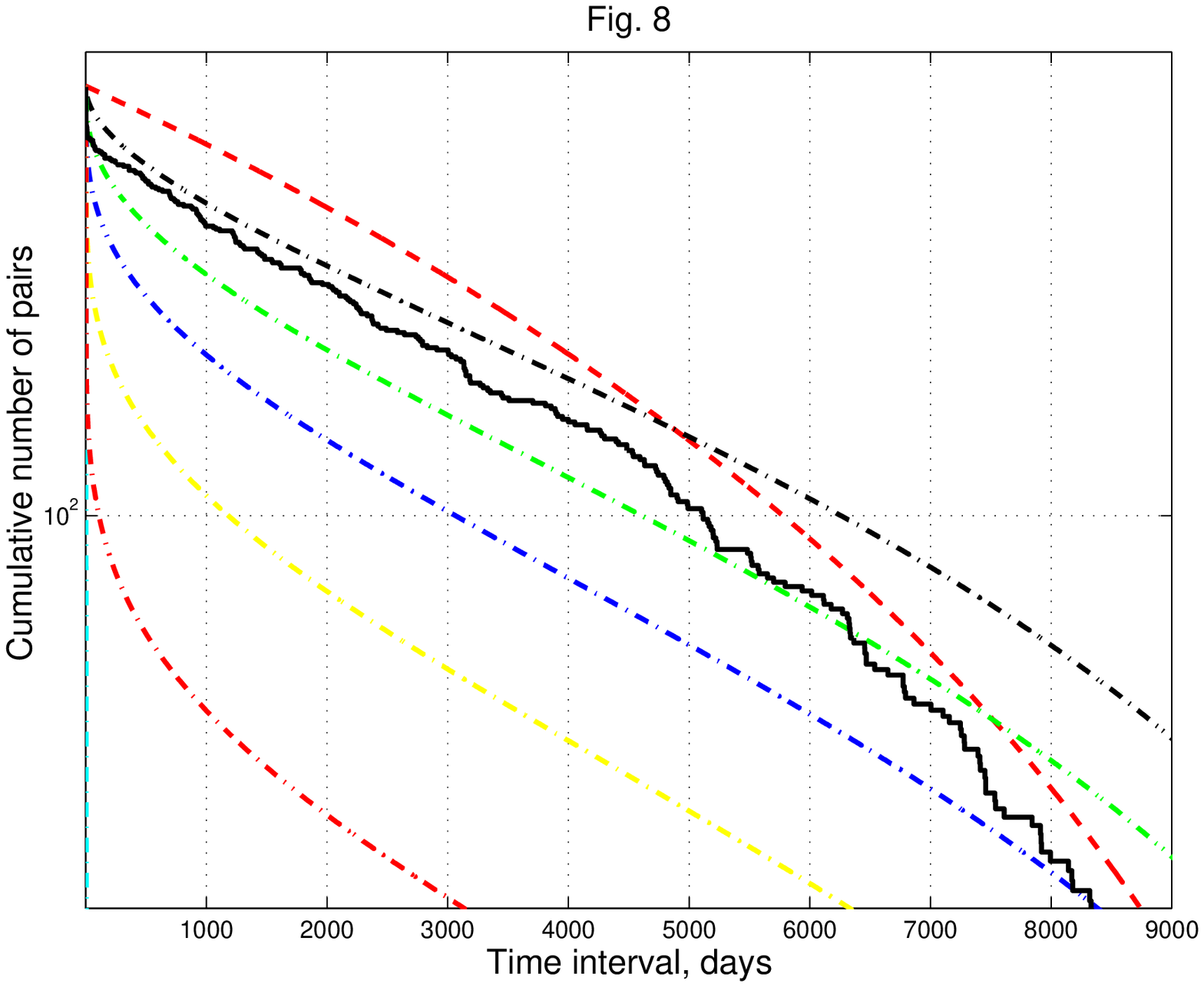}
% #fig09 TIMED1_650_TF.M P53,p.62
\caption{\label{fig09}
}
\end{center}
Earthquake pair numbers in all zones with rates $
26-45 \times 10^{-20}$~events/m$^2$/s.
Pair number is 333.
Average recurrence time ($\bar t$) for earthquakes is $\bar t
\ = \ 3507 \pm 3242$ days.
For notation see Fig.~\ref{fig08}.
\end{figure}

\begin{figure}
\begin{center}
\includegraphics[width=0.75\textwidth]{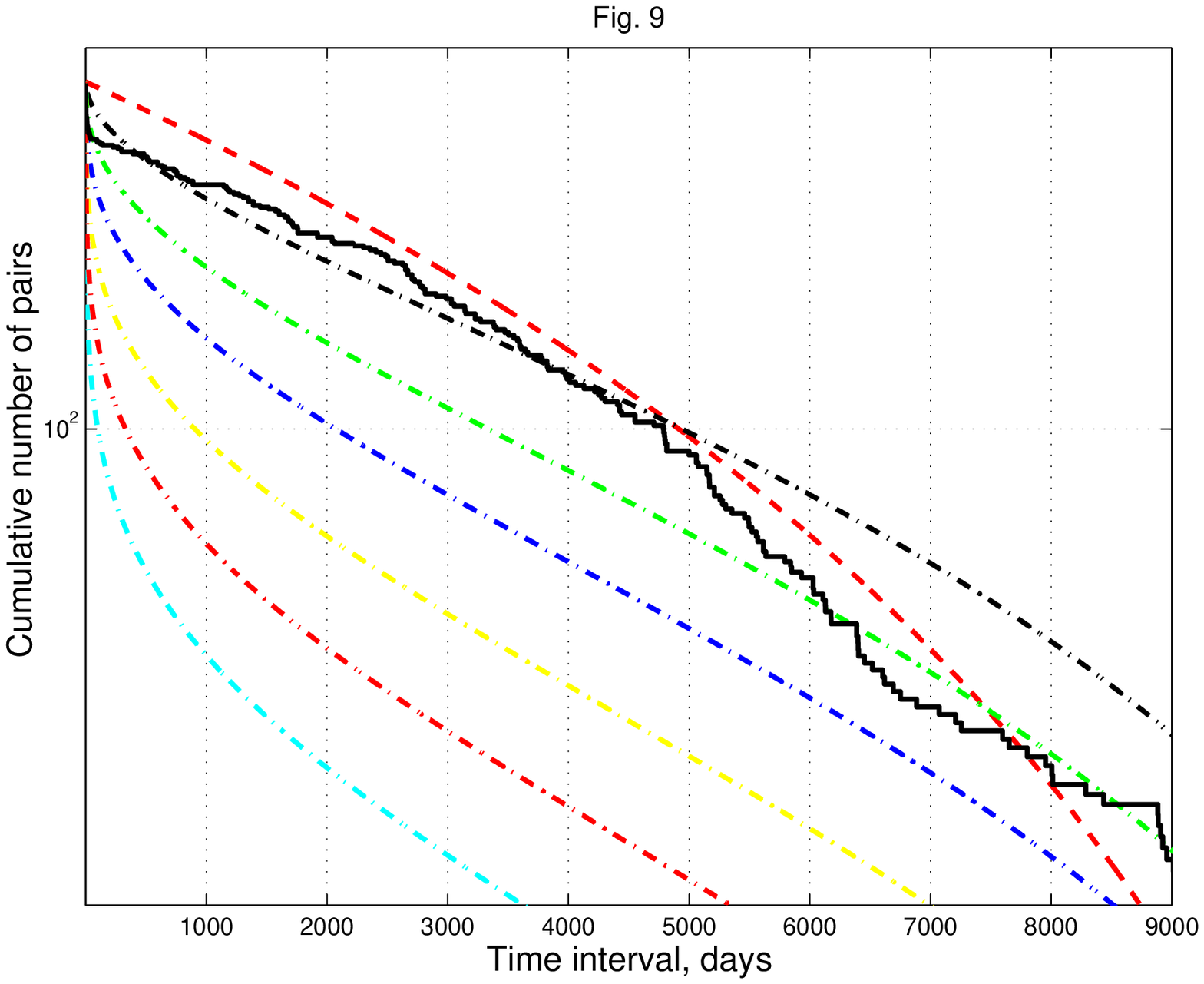}
% #fig10 TIMED1_650_TF.M P53,p.62
\caption{\label{fig10}
}
\end{center}
Earthquake pair numbers in all zones with rates $
\ge 45 \times 10^{-20}$~events/m$^2$/s.
Pair number is 264.
Average recurrence time ($\bar t$) for earthquakes is $\bar t
\ = \ 3947 \pm 3319$ days.
For notation see Fig.~\ref{fig08}.
\end{figure}

\begin{figure}
\begin{center}
\includegraphics[width=0.75\textwidth]{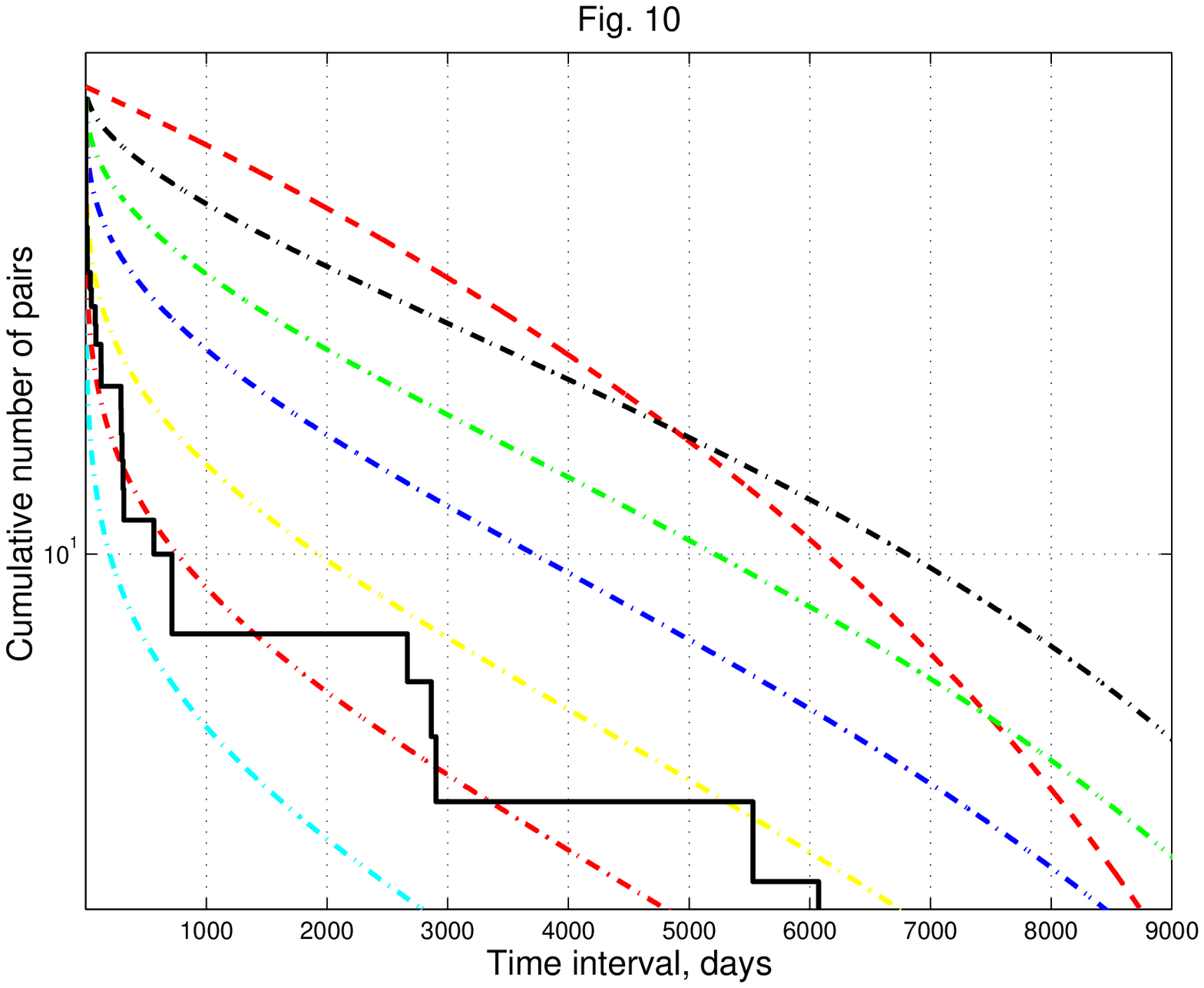}
% #fig11 BIRD_ZONES.FOR FPSM2L.FOR BIRD_TECT2.FOR TIMED1_650_TF_AC.M P53,p.59a
\caption{\label{fig11}
}
\end{center}
Earthquake pair numbers in active continental zones.
Pair number is 37.
Average recurrence time ($\bar t$) for earthquakes is $\bar t
\ = \ 1304 \pm 2578$ days.
For notation see Fig.~\ref{fig08}.
\end{figure}

In Figs.~\ref{fig08}-\ref{fig11} the temporal distribution of
inter-earthquake times for $m \ge 6.5$ event pairs is shown as
it depends on earthquake rate determined by Bird {\it et al.}\
(2010).
Several approximations for pair intervals are also given: the
Poisson distribution of earthquakes (\ref{Eq20}) in the time
span 1976-2010 and a few power-law interval (\ref{Eq21})
dependencies.

Obviously distribution curves consist of two parts: for small
time intervals, they follow a power-law and for larger
intervals the distribution is parallel to the Poisson rate.
The transition from one behavior to another occurs sooner for
zones with a higher tectonic deformation rate: in
Fig.~\ref{fig08} the transition is observed for the time of
about 6,000 days; in Fig.~\ref{fig09} it is about 4,000-5,000
days, and in Fig.~\ref{fig10} it is less than 3,000 days.

Fig.~\ref{fig11} shows the time interval distribution for
active continental areas.
In this case a visual inspection suggests that the best
approximation is the power-law; no transition to the Poisson
rate is observable.
This absence can be explained by a low deformation rate in
these zones (see Fig.~\ref{fig07}).
As was observed for many aftershock sequences in continental
and slowly deforming areas, aftershock sequences continue
according to Omori's law for decades and even centuries
(Utsu {\it et al.}, 1995; Stein and Liu, 2009; Ebel, 2009).
Here the span of 34 years covered by the GCMT catalog
is likely insufficient to demonstrate the transition from an
aftershock sequence to a background, Poisson rate.

We compute an average recurrence time ($\bar t$) for
earthquakes in Figs.~\ref{fig08}-\ref{fig11}.
The smallest value for $\bar t$ is observed for distributions
with a significant component of the power-law:
Figs.~\ref{fig08} and \ref{fig11}.
We also calculate the coefficient of variation ($C_v$) of
earthquake inter-occurrence time as a ratio of the standard
deviation ($\sigma_t$) to the average time $\bar t$ (Kagan and
Jackson, 1991).
A completely random Poisson occurrence has the coefficient of
variation equal to one, whereas quasi-periodicity yields a
coefficient of less than one.
It yields the coefficient larger than one for clustered
earthquakes.
Although the $C_v$ estimates are biased when determined at a
relatively short catalog time span, their mutual relations are
indicative of occurrence patterns.
For Figs.~\ref{fig08}-\ref{fig11}, the $C_v$-values are 1.10,
0.925, 0.841, and 1.977, respectively.
These evaluations again suggest that earthquakes in areas with
a smaller tectonic rate become more clustered, and their
Poisson component is diminished.

\subsection{Stochastic branching processes
}
\label{observ1}

As we mentioned above, the mainshock is often followed by
strong aftershocks and those are clearly accompanied by their
own sequence of events, and so on.
The patterns of multiple clustering have been described by
stochastic branching processes (Hawkes and Adamopoulos, 1973;
Kagan, 1991; Ogata, 1998).
In this model, seismicity is approximated by a Poisson cluster
process, in which clusters or sequences of earthquakes are
statistically independent, although individual events in
the cluster are triggered.
However, Br\'emaud and Massouli\'e (2001) proposed a model in
which all events belong to one branching cluster.

Kagan {\it et al.}\ (2010) applied the Critical Branching
Models (CBM) to analyze earthquake occurrences statistically
in several global and regional earthquake catalogs.
Time intervals between earthquakes within a cluster are
assumed to be distributed according to a power-law
\be
\psi_{\Delta t} \, (\Delta t) \ = \ \theta \, t_M^\theta \,
( \Delta t ) ^{-1-\theta}, \quad \Delta t \ge t_M \, .
\label{Eq22}
\ee
This is similar to Omori's law.
The parameter $\theta$ is an `earthquake memory' factor, $t_M$
is the coda duration time of an earthquake with seismic moment
$M_i$, and $\mu$ is the branching (productivity) coefficient
(Kagan {\it et al.}, 2010, Eq.~9).

\begin{figure}
\begin{center}
\includegraphics[width=0.75\textwidth]{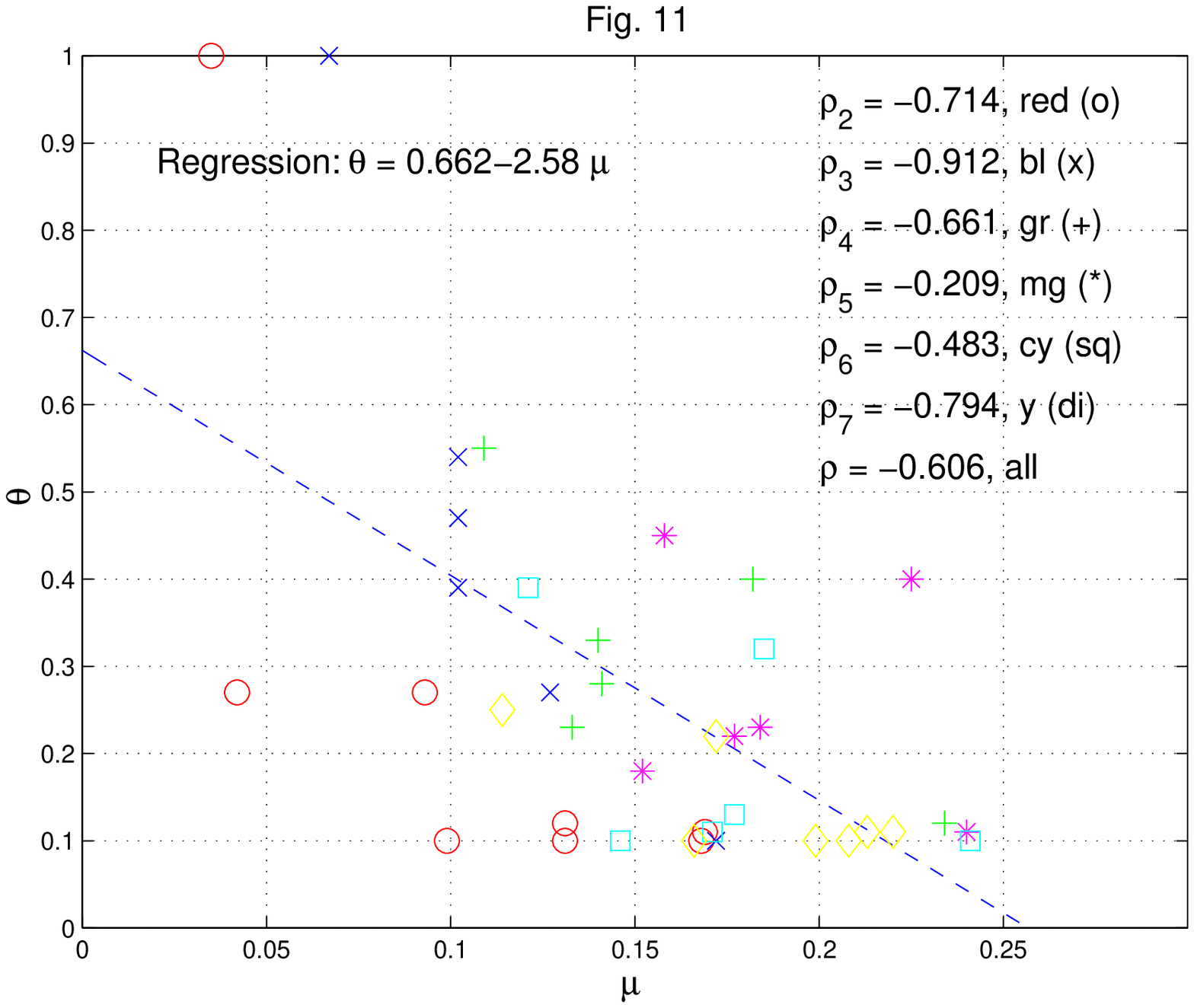}
% #fig12 MU_THETA_CORREL.M P53,p.19a
\caption{\label{fig12}
}
\end{center}
Correlation between maximum likelihood estimates of $\mu$ and
$\theta$ parameters in several earthquake catalogs (Kagan {\it
et al.}, 2010).
The subscripts $i$ at correlation coefficients ($\rho_i$)
point to the table number in Kagan {\it et al.}\ (2010).
\vskip .1in
\end{figure}

During a likelihood search the $\theta$-values have been
restricted within the interval $0.1 \le \theta \le 1.0$;
smaller or larger estimates are inadmissible because of
physical considerations (Kagan, 1991; Kagan {\it et al.},
2010).
Fig.~\ref{fig12} shows a correlation between two parameters
$\theta$ and $\mu$ of the CBM.
Most $\theta$-values are within the interval $0.1 \le \theta
\le 0.5$ and are negatively correlated with the $\mu$
coefficient.

Several determinations of the time decay exponent are carried
out for the ETAS (Epidemic Type Aftershock Sequence) model.
Ogata (1998, Tables~2-3) obtained the $p$-values (equivalent
to our $1+\theta$ coefficient), of the order 1.03--1.14.
Ogata {\it et al.} (2003, Tables~1-2) estimates the
$p$-values to vary within 1.05--1.18.
Ogata and Zhuang's (2006, Tables~2-3) values are 1.02--1.05.
Zhuang {\it et al.} (2005) obtained $p$-values of 1.14--1.15.
Helmstetter {\it et al.}\ (2006, Table~1) used a branching
model similar to the ETAS and calculated $p = 1.18-1.20$.

\begin{figure}
\begin{center}
\includegraphics[width=0.75\textwidth]{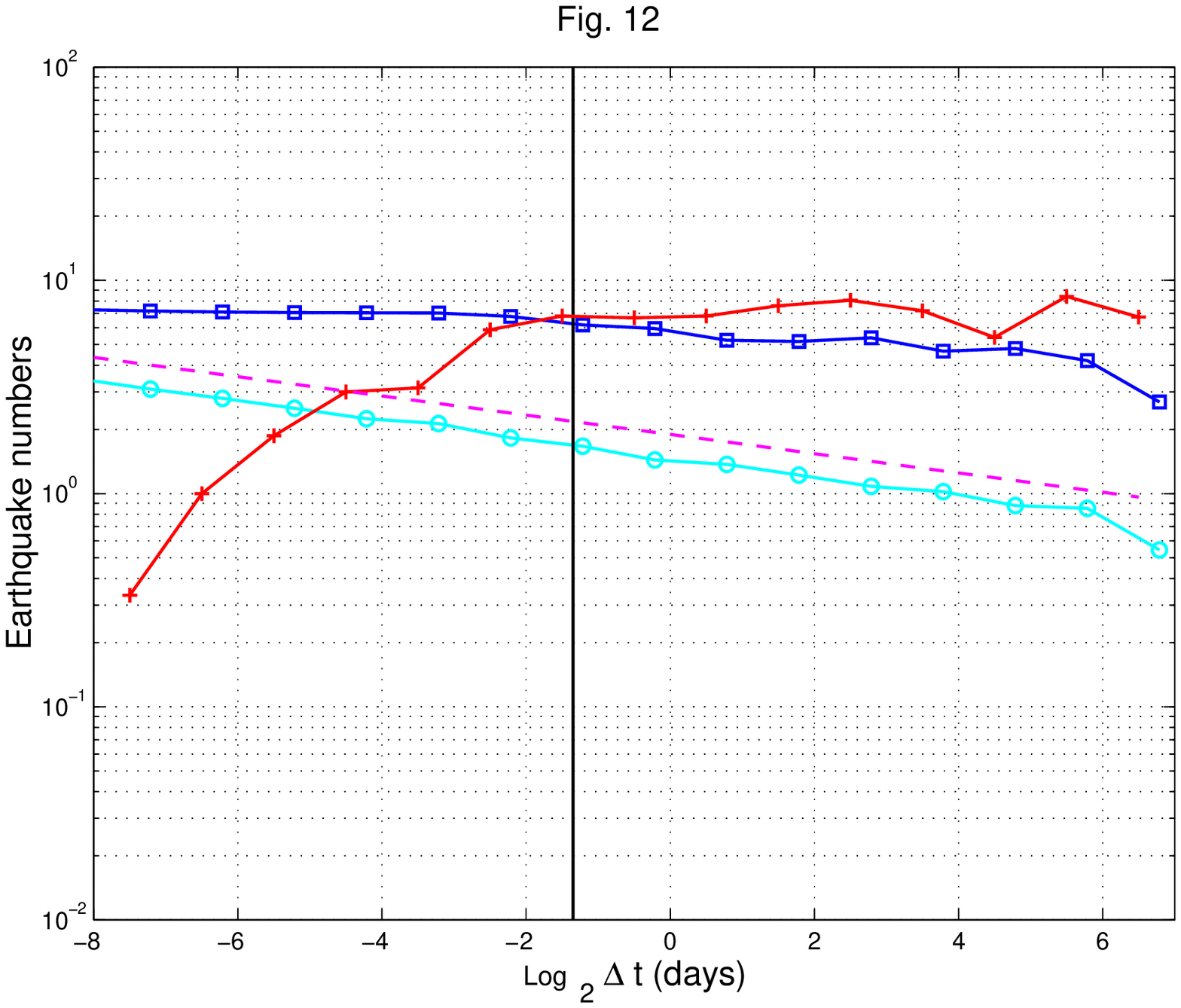}
% #fig13 AFTI5A4.m P53,pp.20,21
\caption{\label{fig13}
}
\end{center}
The average aftershock numbers from the PDE catalog in
logarithmic time intervals following $m \ge 8.0$ GCMT
earthquakes during 1977-2003.
Line with plus signs shows $m_b~\ge~4.9$ aftershocks; dashed
lines are theoretical estimates for the first generation
aftershocks.
Two curves display results of earthquake catalog simulation:
line with circles shows first generation aftershocks, line
with squares indicates total aftershock numbers.
\end{figure}

To illustrate our fit of the temporal distribution of dependent
earthquakes, average numbers of aftershocks following 15 $m
\ge 8.0$ GCMT earthquakes are displayed in Fig.~\ref{fig13}
(similar to Fig.~13 in Kagan, 2004).
We use the time period 1977-2003, so that all large
earthquakes are approximately the same size ($8.45 \ge m
\ge 8.0$, i.e., excluding the 2004 Sumatra and its
aftershocks).
Since the GCMT catalog has relatively few dependent events, we
selected aftershocks from the the U.S.\
Geological Survey (2010) PDE (Preliminary Determinations of
Epicenters) catalog.
The aftershock rate in the diagram is approximately constant
above our estimate of the coda duration $t_M$.
For the logarithmic intervals, this corresponds to the
standard form of the Omori law: the aftershock number $n_a$ is
proportional to $1/t$.
For the smaller time intervals, the aftershock numbers decline
when compared to the Omori law prediction ($1/t$).
This decline is caused by several factors, the interference of
mainshock coda waves being the most influential (Kagan, 2004).
The decline is faster for weaker events ({\it ibid.}).
% Because our forecast is calculated once per day, these
% immediate aftershocks usually die out before the
% forecast is updated.

For theoretical estimates, we used the values of parameters
obtained during the likelihood function search (Kagan {\it et
al.}, 2010, Table~4) obtained for the full PDE catalog, $m_t =
5.0$: the branching coefficient $\mu = 0.141$, the parent
productivity exponent $a_0=0.63$ ($a_0=\kappa \times 1.5$),
and the time decay exponent $\theta = 0.28$.
Theoretical estimates in Fig.~\ref{fig13} seem to be
reasonably good at forecasting time intervals on the order of
one day.
For larger intervals, the expected numbers decrease as $n_a
\sim (\Delta t)^{-1.15}$: this is stronger than the regular
Omori law would predict.
As we suggested in Section~\ref{analys1} the Omori law assumes
that all aftershocks are direct consequences of a mainshock,
whereas a branching model regards any earthquake as a possible
progenitor of later events.
Thus, later aftershocks are the combined offspring of a
mainshock and all consequent earthquakes.
With increase in time, the difference between Omori's law and
the CBM predictions would increase as well.
Marsan and Lenglin\'e (2008) show that in California catalogs
due to cascading aftershock rates for direct and secondary
triggering differ by a factor of 10 to about 50\%.
By numerical simulations of the ETAS model, Felzer {\it et
al.}, (2002) and Helmstetter and Sornette (2003) estimated
that a substantial majority of aftershocks are indirectly
triggered by the mainshock.

The $p$-parameter in Omori's law is often assumed to be 1.0.
This $p$-value implies that the total aftershock number
approaches infinity as the duration of the aftershock sequence
increases.
Figs.~\ref{fig03} and \ref{fig04}, however, suggest that in the
presence of tectonic loading, the time exponent value should
increase at longer time periods.
Statistical determination of the exponent can usually be made
for only short periods; thus, we do not yet have a good
estimate of the $p$-value for the tail of an aftershock
sequence.

\section{Earthquake statistics and plate-tectonic
deformation }
\label{tecto}

\subsection{Stationary earthquake rate due to tectonic
deformation }
\label{analys3}

The tapered Gutenberg-Richter (TGR) relation (Kagan, 2002b)
has an exponential taper applied to the number of events with
a large seismic moment.
Its survivor function ($1 \ -$ cumulative distribution)
for the scalar seismic moment $M$ is
\be
F (M) \ = \ (M_t/M)^{\beta} \,
\exp \left ( {{M_t \, - \, M} \over M_{c} } \right ) \quad
{\rm for} \quad M_t \le M < \infty \, .
\label{Eq13}
\ee
Here $M_{c}$ is the parameter controlling the distribution in
the upper ranges of $M$ (`the corner moment'), $M_t$ is the
moment threshold: the smallest moment above which
the catalog can be considered to be complete; $\beta$ is
the index parameter of the distribution; $ \, \beta = {2 \over
3} b \, $, $b$ is a familiar $b$-value of the
Gutenberg-Richter distribution (G-R, Gutenberg and Richter,
1944).

By evaluating the first moment of the distribution
(\ref{Eq13}), we can obtain a theoretical estimate of the
seismic moment flux (Kagan, 2002b)
\be
\dot M_s \ = \ { { \alpha_0 \, M_0^\beta } \over {1 -
\beta }} \, M_{c}^{1 - \beta} \, \Gamma (2 - \beta) \,
\exp \, (M_0/M_{c}) \, ,
\label{Eq14}
\ee
where $\Gamma $ is a gamma function and $\alpha_0$ is the
seismic activity level (occurrence rate) for earthquakes with
moment $M_0$ and greater.

We compare the seismic moment rate with the tectonic moment
rate ($\dot M_T$):
\be
\dot M_T  \ = \ \mu \, W \int\limits_{ \ \ A} \int | \dot
\epsilon | \, dA \ = \ \dot M_s \, / \, \chi \, ,
\label{Eq15}
\ee
where $\chi$ is the seismic coupling (or seismic efficiency)
coefficient, $\mu$ is the elastic shear modulus, $W$ is the
seismogenic width of the lithosphere, $\dot \epsilon$ is
the average horizontal strain rate, and $A$ is the area under
consideration.
At present, some variables in the equation cannot be evaluated
with great accuracy; to overcome this difficulty we calculate
a product of these variables: the `effective width' of
seismogenic zone $W_e$ or coupled seismogenic thickness
(Bird and Kagan, 2004):
\be
W_e = W \, \chi \, .
\label{Eq16}
\ee
Bird and Kagan (2004, Eq.~11) propose another, more exact
formula for calculating the tectonic moment rate appropriate
to a plate boundary zone.

In regions of high seismicity, instead of
Eqs.~\ref{Eq14}--\ref{Eq16} we can use measured long-term
seismic activity to infer the earthquake rate by extrapolating
(\ref{Eq13}) to any moment level (see Bird and Liu, 2007,
Eqs.~4 and 5).
Bird {\it et al.}\ (2010) presented an algorithm and tables
for a long-term world-wide forecast of shallow seismicity
based on the Global Strain Rate Map (GSRM) by Kreemer {\it et
al.}\ (2003).
% The forecast is a set of global maps represented by gridded
% values ($89.95^\circ$S to $89.95^\circ$N, $0.05^\circ$E to
% $360.05^\circ$E, in $0.1^\circ$ steps) of the forecast rate
% for shallow earthquake epicenters per square meter per second
% (including aftershocks) above threshold.
% Two tables with representative thresholds ($m_t = 5.66$ and
% $8.00$) are stored as an electronic supplement ({\it ibid.}).
Because GSRM does not estimate tectonic strain-rates of stable
plate interiors, a simple empirical-averaging method has been
used.
Thus, the seismicity in plate interiors is represented by a
uniform rate.

Since the seismicity level in plate interiors may vary by
orders of magnitude, the uniform rate may strongly under-
or over-estimate the seismicity rate.
Therefore, we apply Eqs.~\ref{Eq14}--\ref{Eq16} to
evaluate first the tectonic moment rate and then a long-term
forecast for these regions
\be
\alpha_0 \ = \ { { \dot M_T \, \chi \, (1 -
\beta ) \, \exp \, (-M_0/M_{c}) }
\over  { \, M_0^\beta \, M_{c}^{1 - \beta} \,
\Gamma (2 - \beta) \, } } \, .
\label{Eq17}
\ee
By calculating $\alpha_0 $ for a particular choice of $M_0$,
we may re-normalize Eq.~\ref{Eq13} and obtain earthquake size
distribution for any region with a known strain rate and
corner moment.

\subsection{Length of aftershock zone
}
\label{analys4}

Eq.~\ref{Eq17} above can be used to evaluate the background
seismicity level for an aftershock zone.
Such a calculation would use the area of an earthquake focal
zone.
This area can be estimated by a dimension of aftershock zone
for each event.
Kagan (2002a) evaluated how the aftershock zone size for
mainshocks $m \ge 7.0$ depends upon on the earthquake
magnitude by approximating aftershock epicenter maps through a
two-dimensional Gaussian distribution.
The major ellipse axis is taken as a quantitative measure of
the mainshock focal zone size.

In Fig.~\ref{fig05} we display the regression curves for
GCMT/PDE earthquakes: all earthquakes for three choices of
focal mechanisms (updated Fig.~6a by Kagan, 2002a).
In regression curves we use $m = 8.25$ as a reference point.
For example, for the quadratic regression
\be
L \ = \ \log_{10} \, \ell \ = \ a_0 \, + \, a_1 \, (m -
8.25) \, + \, a_2 \, (m - 8.25)^2
\, ,
\label{Eq18}
\ee
where $\ell$ is the length of the aftershock zone in km.
For the linear regression we set $a_2 = 0$.
Fig.~\ref{fig06} displays the regression in a similar format
for active continental tectonic zones (Kagan {\it et al.},
2010).

Table~\ref{Table1} summarizes the results of regression
analysis for all global earthquakes, as well as for events in
subduction zones (trenches) and on active continental zones.
Earthquakes are also subdivided by their focal mechanism.
Other tectonic zones lack a sufficient number of $m \ge 7.0$
mainshocks to carry out this statistical analysis.

The following conclusions can be made from Table~\ref{Table1}:
(a) aftershock zones exhibit similar scaling;
(b) zone length ($\ell$) on average is proportional to moment
$M^{1/3}$; and
(c) the value of $a_0$ parameter (zone length for the $m8.25$
earthquake) is close to $10^{2.5}$ (316) km for all cases.
Normal earthquakes (rows 5-6) exhibit slightly different
scaling; zone length ($\ell$) for the linear regression is
proportional to moment $M^{1/2.8}$.
Scaling for strike-slip earthquakes (rows 7-8) also differs a
little from average: zone length ($\ell$) is proportional to
moment $M^{1/3.5}$.
However, the earthquake numbers in these subsets are small,
thus it is possible that these variations are due to random
fluctuations.

Only three subsets show a substantial nonlinearity:
(a) trenches with strike-slip focal mechanisms (rows 15-16);
(b) continents with all focal mechanisms (rows 17-18); and
(c) continents with strike-slip focal mechanisms (rows 21-22).
However, the earthquake numbers are small in all these plots,
and although (b) and (c) aftershocks display zone lengths which
increase strongly for the largest earthquakes (the feature
often quoted in other studies of length scaling -- see, for
instance, Wells, and Coppersmith, 1994, and subsequent
publications citing that paper), (a) earthquakes exhibit an
opposite behavior.

In all diagrams the standard errors ($\sigma$) are almost the
same for the linear and quadratic regression.
The maximum errors ($\epsilon_{\rm max}$) follow the same
pattern.
This pattern suggest that linear regression is sufficient to
approximate the data.
Although the quadratic regression fit yields no statistically
significant improvement in almost any diagram, the sign of the
quadratic correction term is negative for most cases.
The negative value of the $a_2$ regression coefficient means
that increase in the aftershock zone length is weaker for the
largest earthquakes.
This feature contradicts those often quoted in other studies
of length scaling (see Kagan, 2002a for details).
Thus, the slope of the regression curve is either stable or
decreases at the high magnitude end.
No saturation effect for large earthquakes occurs in the data.
Results in Table~\ref{Table1} imply that the major ellipse
axis $a$ (length) of an earthquake focal zone can be
approximated
by
\be
a \ = \ 316 \times 10^{(m - 8.25)/2} \ {\rm km}
\, .
\label{Eq18a}
\ee

We conclude that earthquake rupture length is proportional to
the cube root of moment, which implies that width and slip
should scale the same way.
Otherwise, one of them increases less strongly with moment and
the other more strongly. For either that would pose the
problem of ``inverse saturation."

We assume that the majority of aftershocks are concentrated
within an ellipse having 2-$a$ major axis.
The probability that a point lies inside a 2-$a$ ellipse
is shown in Eq.~5 by Kagan (2002a).
If we know the length of an earthquake focal zone, we can
calculate its area.
We assume, for example, that the ratio of the major ellipse
axis to the minor axis is 1/4; then area $S$ of the focal zone
is
\be
S \ = \ \pi a^2
\, .
\label{Eq18b}
\ee

\subsection{Example: New Madrid earthquake sequence of 1811-12
}
\label{observ2}

To illustrate the arguments and results of the previous
sections, we calculate seismicity parameters of the New Madrid
earthquake sequence (1811-12) and its consequences.
There is substantial literature on this sequence (Stein and
Liu, 2009; Calais {\it et al.}, 2010, and references therein).

Three or four large earthquakes with magnitudes on the order
of 7.5--8.0 occurred over a few months of 1811-12 in the New
Madrid area; aftershocks of these events are still registered.
As an illustration, we would assume that only one $m8$ event
occurred at that time.
If in reality earthquakes were smaller than such an event,
their total focal zone and combined aftershock sequence at the
present time would be equivalent to about one $m8$ mainshock.

\begin{figure}
\begin{center}
\includegraphics[width=0.75\textwidth]{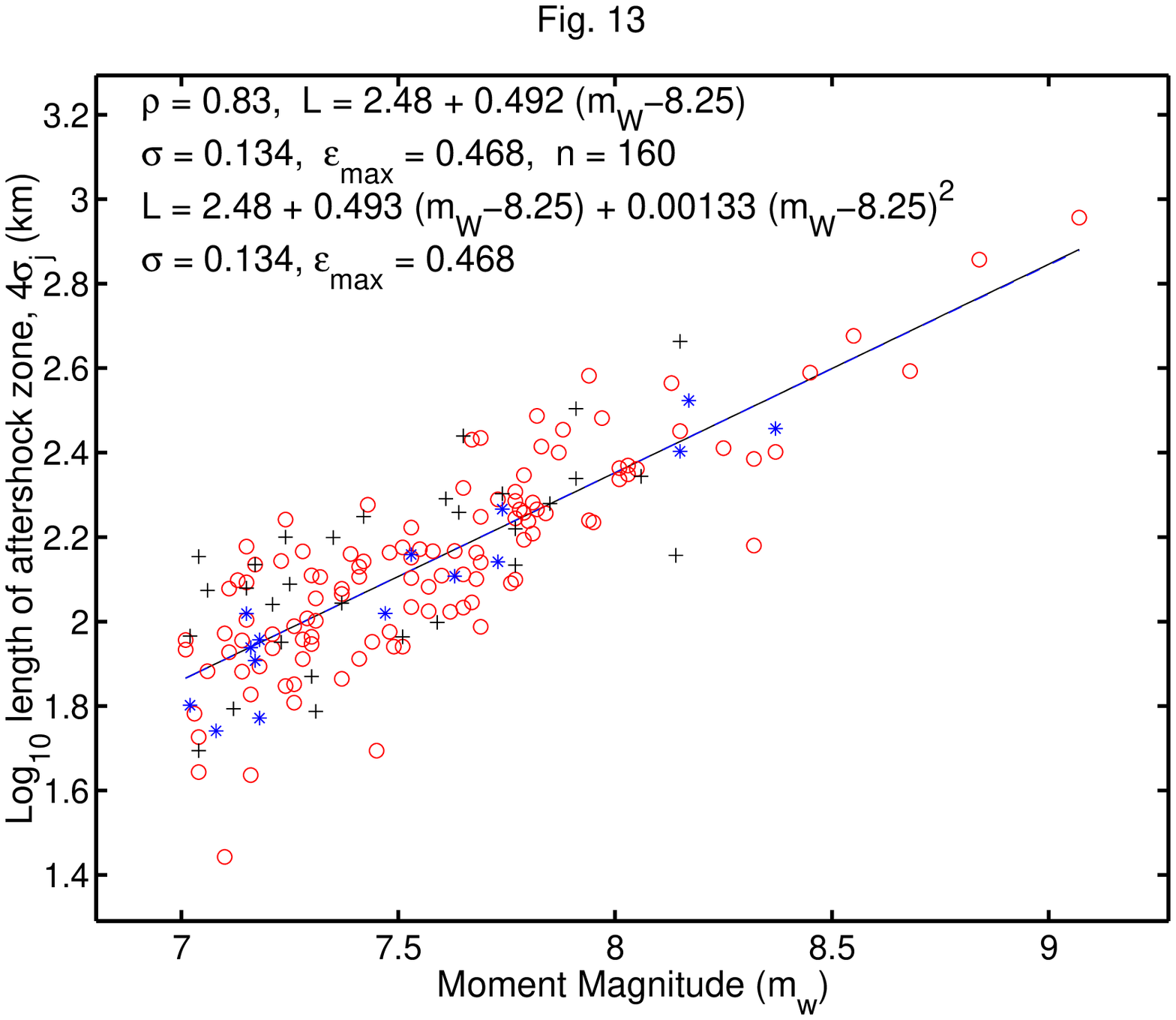}
% #fig05 SCAL1A1.m P53,p.25
\caption{\label{fig05}
}
\end{center}
Plot of log aftershock zone length ($L$) against moment
magnitude ($m$).
% Magnitude values are shifted in formulas shown in the plot
% ($m_r = m - 8.25$).
Rupture length is determined using a 1-day aftershock pattern.
Values of the correlation coefficient ($\rho$),
coefficients for linear (dashed line) and quadratic (solid
line) regression, standard ($\sigma$) and maximum
($\epsilon_{\rm max}$) errors, and the total number ($n$) of
aftershock sequences are shown in the diagram.
% The dashed line is the linear regression, the solid line is a
% quadratic approximation.
\hfil\break
Circles -- thrust mainshocks;
\hfil\break
Stars -- normal mainshocks;
\hfil\break
Pluses -- strike-slip mainshocks.
\vskip .1in
\end{figure}

\begin{figure}
\begin{center}
\includegraphics[width=0.75\textwidth]{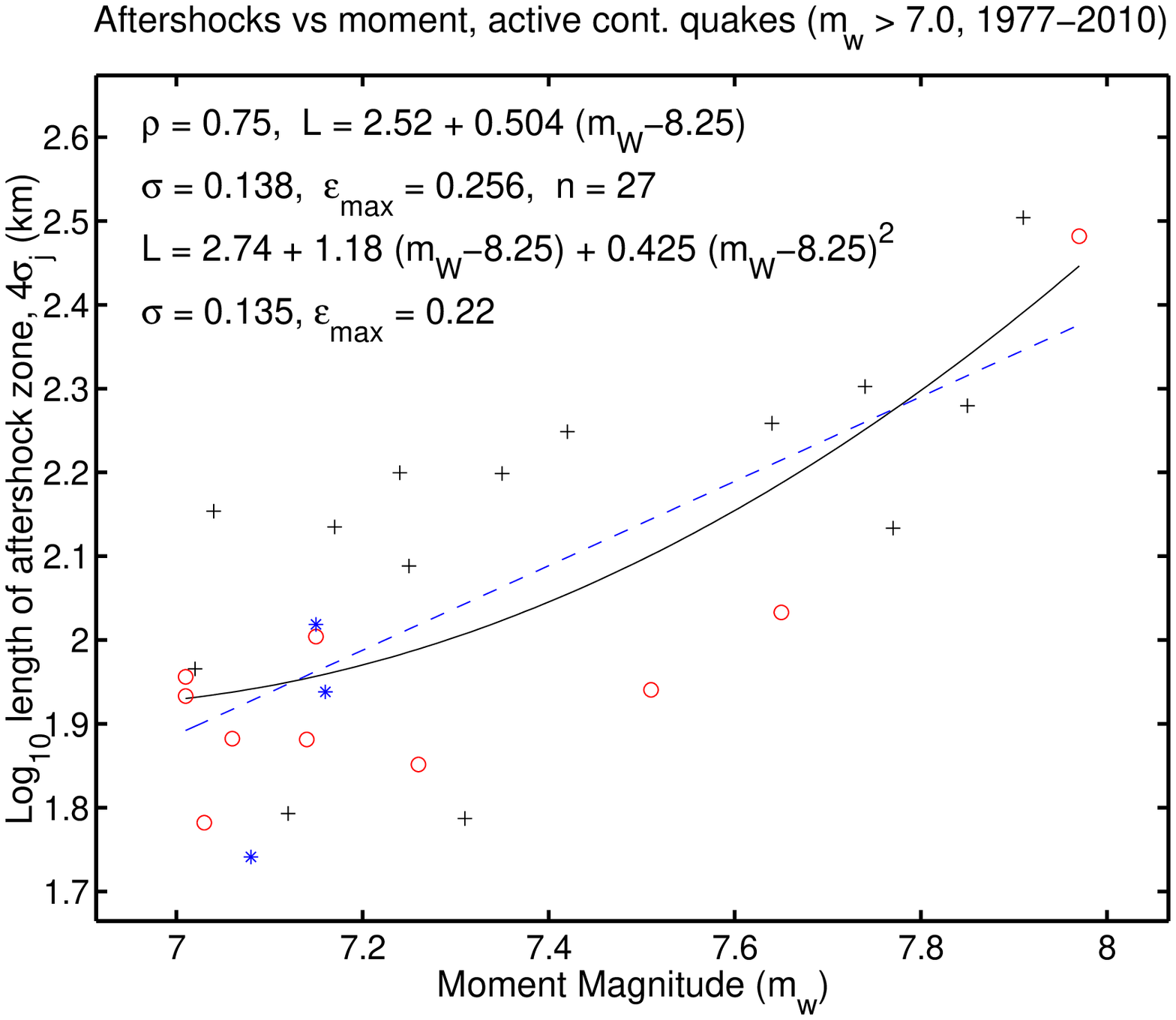}
% #fig06 SCAL1A1.m P53,p.25
\caption{\label{fig06}
}
\end{center}
Plot of log aftershock zone length ($L$) against moment
magnitude ($m$) for earthquakes in active continental zones
(Kagan {\it et al.}, 2010).
For notation see Fig.~\ref{fig05}.
\vskip .1in
\end{figure}

The size of the focal zone can be evaluated by using
regression equations in Figs.~\ref{fig05} and \ref{fig06}.
The first plot contains many earthquakes but most of these
events are in subduction zones.
The second diagram uses earthquakes in active continental
zones; a focal size of these earthquakes is likely to resemble
the New Madrid area which can be classified as plate-interior
(Kagan {\it et al.}, 2010).
Too few large earthquakes are available within plate-interior
to obtain their features.
The difference between regression parameters in
Figs.~\ref{fig05} and \ref{fig06} is small; therefore the
size of earthquake focal zones either does not change in
various tectonic zones or changes slightly.

For an $m8$ earthquake, calculations yield 227~km and 259~km
as the length of the focal zone, defined as the $4 \sigma$
major axis of an ellipse comprising a majority of aftershocks
(Kagan, 2002a).
The linear regression is used in both cases: the former value
corresponds to Fig.~\ref{fig05} and the latter to
Fig.~\ref{fig06}.
The two estimates are similar and roughly correspond to
the size of the present aftershock zone, as shown, for
example, in Calais {\it et al.}\ (2010).

To calculate the surface area of an aftershock zone, we assume
that the minor axis of the ellipse is 1/4 of the major axis,
taken as 240~km.
Then we obtain the $m8$ earthquake focal area as 11300~km$^2$.
Taking the average strain rate as $\dot \epsilon = 10^{-9}$
(Calais {\it et al.}, 2006), we compute the tectonic moment
rate (\ref{Eq15}): $3.4 \times 10^{15}$~Nm/year.
Assuming that 50\% of the tectonic rate is released
seismically (Bird and Kagan, 2004), we obtain the background
rate $\alpha_0 = 1.27 \times 10^{-3}$ $m \ge 5$ earthquakes
per year [we take in (\ref{Eq17}) $M_c = 10^{21}$~Nm, $W
= 20$~km, $\beta = 2/3$; then $\Gamma (4/3) = 0.893$].
We use (\ref{Eq13}) to calculate the recurrence time for an
$m \ge 8$ earthquake in the focal zone of the New Madrid
events: more than two million years.
In this computation any $m8$ earthquake with an epicenter or
centroid in the focal zone counts:
in Eq.~(\ref{Eq17}) we do not request that the entire rupture
of such an earthquake be contained in the zone.
The recurrence time is an average value; even for events
as large as $m \ge 8$ the earthquake occurrence is clustered
(see Section~\ref{pair}).
Thus, a new large earthquake can follow after a relatively
short time period, as exemplified by the 1811-12 New Madrid
sequence.

We would like to calculate the duration of an aftershock
sequence up until the aftershock rate decays to the background
level.
The results in Fig.~\ref{fig13} can be applied to this
purpose.
However, we need to make a correction for the mainshock
and aftershock magnitudes ($m8$ instead of $m8.15$ and $m \ge
5.0$ instead of $m_b \ge 4.9$ in the plot, respectively).
In the diagram the aftershock rate per one interval (the
intervals increase consequently by a factor of two) is 7
events.
This translates into 4.55 $m \ge 5$ events per
interval for our choice of magnitudes (Kagan {\it at al.},
2010).
After comparing the background and aftershock rates (we take
4.55 $m \ge 5$ aftershocks per the first day, decaying
according to Omori's law, with $1/t$ rate with time), we
discover that the aftershock sequence would reach the
background rate in about 3,600~years.
This duration estimate agree roughly with Stein and Liu's
(2009) value.
In these calculations, we presume that no independent large
earthquake clusters would occur during the aftershock
sequence.
The possible occurrence of spontaneous events makes any
evaluation of aftershock sequence duration largely
approximate.

Stein and Liu (2009) obtained aftershock duration values for
several sequences using Eq.~14 from Dieterich (1994).
This equation employs parameters whose values for actual
earthquake focal zones are not known.
Generally, the parameters have been back adjusted based on
the statistics of earthquake occurrence.
This may explain apparently reasonable fit of
Dieterich's formula to aftershock sequences.

In contrast, we obtain the aftershock sequence duration by
extending the tapered Gutenberg-Richter and Omori's laws
and using their well-known properties and measured geometrical
features of tectonic deformation.
Moreover, according to Stein and Liu (2009, Fig.~1c) the New
Madrid aftershock rate for the last 50 years was about 0.5 $m
\ge 4$ events per year.
Computations based on Omori's law similar to those shown
above, yield the rate 175~years after the mainshock occurrence
of about $0.26$ $m \ge 4$ events per year.
This number is close to that shown above.

\section{Discussion }
\label{disc}

Two classical statistical earthquake distributions largely
governed our approach to analyzing seismicity: Omori's law
and the Gutenberg-Richter relation.
As explained above, recent developments in earthquake size
statistics considerably improved our understanding of
earthquake occurrence and could lead to significantly better
estimates of seismic hazard.
For the G-R law the earthquake temporal distribution is mostly
irrelevant, since size distribution of clustered events is
largely independent of their history.
Similar progress in understanding earthquake time statistics
is much more difficult to achieve.
We cannot ignore spatial variables and the available data are
not as extensive, so the task is more complex.
Only by applying rigorous methods, by analyzing carefully
systematic and random effects, and by critical testing of
models and hypotheses we will be able to advance in solving
this problem.

In previous sections we derived the time distribution for
earthquake occurrence; the distribution is shown to be
controlled by power-laws.
How can the parameters of these distributions be determined?
If one excludes the interiors of plates, the tectonic
deformation rate $V$ is reasonably well known for plate
boundaries and for active continents (Kagan {\it et al.},
2010; Bird {\it et al.}, 2010).
The diffusion rate $D$ is presently unknown.
If we could obtain the earthquake temporal distribution as
shown in Figs.~\ref{fig01} and \ref{fig02}, the $D$ evaluation
would be easy.
These distributions are derived for a particular area within
an earthquake fault zone.
However, if the stress in the focal zone of an earthquake is
distributed according to the power-law with an exponent $\psi$
(see Eqs.~\ref{Eq02} and \ref{Eq04}), the problem becomes
more acute.

Figs.~\ref{fig03} and \ref{fig04} suggest that the
distribution temporal behavior changes when $V = \sqrt D$.
This change relates to the first generation of offspring.
Thus, we should not be able to see it in regular Omori plots
which combine many generations of aftershocks.
The inversion of earthquake occurrence parameters based on
stochastic branching processes yields needed first generation
effects.
However, in present models (both CBM and ETAS) as
discussed in Section~\ref{observ1}, temporal dependence is
parameterized by just one exponent.
These models should in principle demonstrate changes in the
temporal pattern, if more complicated temporal function is
applied.
However, results from statistical analysis are very
uncertain even for one-parameter time decay.
Given the contemporary quantity and quality of earthquake
catalogs, it is unlikely more complicated models would be
effective in resolving this issue.

Perhaps new laboratory experiments (Zaiser, 2006) may help
solve the problem of diffusion rate evaluation, but it is not
clear whether such measurements are possible.
The acoustic emission event rate exhibits fore- and aftershock
sequences associated with dynamic failure of the test
specimen (see, for example, Ojala {\it et al.}, 2004).
These and similar tests can be used to infer the dependence of
the Omori law parameters on spatial scale and stress
diffusion rate.

Results from statistical analysis of earthquake occurrence
in our previous publications (Kagan, 2002b; Bird
and Kagan, 2004; Kagan {\it et al.}, 2010), as well as the
results reported above, suggest that the earthquake process in
all tectonic provinces can be described by the same model.
We advocate the Poisson cluster process with clusters
controlled by a critical branching process and a power-law
time dependence.
Combined with the earthquake size distribution approximation
by the TGR, such a model allows mathematically forecast
of a spatially variable, time-independent (long-term) earthquake
rate.
It will optimally smooth the seismicity record (Kagan and
Jackson, 2010) or translate the plate-tectonic and geodetic
rate into a seismic hazard estimate (Bird and Liu, 2007; Bird
{\it et al.}, 2010).

A short-term forecast can be performed by using the temporal
properties of earthquake clusters, an extrapolation which uses
a variant of Omori's law to estimate future earthquake rate
(Kagan and Jackson, 2010).
In Section~\ref{observ2} we presented an example of such
calculations.

In conclusion,
this paper suggests a method for calculating
long- and short-term seismicity estimates, based on
theoretical inference about classical, statistical earthquake
distributions: the Omori law and the G-R relation.
Statistical analysis of earthquake occurrence carried out in
our previous papers and in this work make such seismic hazard
evaluation more reliable and accurate.

\subsection* {Acknowledgments
}
\label{Ackn}
I am grateful to Dave Jackson, Paul Davis, and Peter Bird
for useful discussion and suggestions.
Per J\"ogi helped with the {\scaps mathematica} computations.
I thank Kathleen Jackson for significant improvements in the
text.
The author appreciates support from the National Science
Foundation through grant EAR-0711515, as well as from the
Southern California Earthquake Center (SCEC).
SCEC is funded by NSF Cooperative Agreement EAR-0529922
and USGS Cooperative Agreement 07HQAG0008.
Publication 0000, SCEC.

\pagebreak

\centerline { {\sc References} }
\vskip 0.1in
\parskip 1pt
\parindent=1mm
\def\reference{\hangindent=22pt\hangafter=1}

\reference
Abramowitz, M. and Stegun, I.~A., (Eds.), 1972.
{\sl Handbook of Mathematical Functions}, Dover, NY, pp.~1046.

\reference
Bakun, W.\ H., and A.\ G.\ Lindh, 1985.
The Parkfield, California, earthquake prediction experiment,
{\sl Science}, {\bf 229}, 619-624.

\reference
Bakun, W. H., B. Aagaard, B. Dost, W. L. Ellsworth, J. L.
Hardebeck, R. A. Harris, C. Ji, M. J. S. Johnston, J.
Langbein, J. J. Lienkaemper, A. J. Michael, J. R. Murray, R.
M. Nadeau, P. A. Reasenberg, M. S. Reichle, E. A. Roeloffs, A.
Shakal, R. W. Simpson, and F. Waldhauser, 2005.
Implications for prediction and hazard assessment from the
2004 Parkfield earthquake,
{\sl Nature}, {\bf 437}, 969-974.

\reference
Bird, P., and Y. Y. Kagan, 2004.
Plate-tectonic analysis of shallow seismicity: apparent
boundary width, beta, corner magnitude, coupled
lithosphere thickness, and coupling in seven tectonic settings,
{\sl Bull.\ Seismol.\ Soc.\ Amer.}, {\bf 94}(6), 2380-2399
(plus electronic supplement).

\reference
Bird, P., C. Kreemer, and W. E. Holt, 2010.
A long-term forecast of shallow seismicity based on the
Global Strain Rate Map,
{\sl Seismol.\ Res.\ Lett.}, {\bf 81}(2), 184-194
(plus electronic supplement).

\reference
Bird, P., and Z. Liu, 2007.
Seismic hazard inferred from tectonics: California,
{\sl Seism.\ Res.\ Lett.}, {\bf 78}(1), 37-48.

\reference
Br\'emaud, P., and L. Massouli\'e, 2001.
Hawkes branching point processes without ancestors,
{\sl J. Applied Probability}, {\bf 38}(1), 122-135.

\reference
Calais, E., Han, J. Y., DeMets, C., Nocquet, J. M., 2006.
Deformation of the North American plate interior from a decade
of continuous GPS measurements,
{\sl J. Geophys.\ Res.}, {\bf 111}(B6), B06402,
doi:10.1029/2005JB004253.

\reference
Calais, E., Freed, A. M., Van Arsdale, R. \& Stein, S. (2010).
Triggering of New Madrid seismicity by late-Pleistocene
erosion,
{\sl Nature}, {\bf 466}, 608-611

\reference
Chhikara, R., and J. L. Folks, 1989.
{\sl The Inverse Gaussian Distribution},
New York, M. Dekker, 213~pp.

\reference
Dawson, T. E., S. F. McGill, and T. K. Rockwell (2003),
Irregular recurrence of paleoearthquakes along the
central Garlock fault near El Paso Peaks, California,
{\sl J. Geophys.\ Res.}, {\sl 108}(B7), 2356,
doi:10.1029/2001JB001744.

\reference
Dieterich, J., 1994.
A constitutive law for rate of earthquake production and its
application to earthquake clustering,
{\sl J. Geophys.\ Res.}, {\bf 99}, 2601-2618.

\reference
Ebel, J. E., 2009.
Analysis of aftershock and foreshock activity in stable
continental regions: implications for aftershock forecasting
and the hazard of strong earthquakes,
{\sl Seismological Research Letters}, {\bf 80}(6), 1062-1068.

\reference
Ekstr\"om, G., 2007.
Global seismicity: results from systematic waveform analyses,
1976-2005,
in {\sl Treatise on Geophysics}, {\bf 4}(4.16), ed.\ H.
Kanamori, pp.~473-481.

\reference
Ekstr\"om, G., A. M. Dziewonski, N. N. Maternovskaya
and M. Nettles, 2005.
Global seismicity of 2003: Centroid-moment-tensor solutions
for 1087 earthquakes,
{\sl Phys.\ Earth planet.\ Inter.}, {\bf 148}(2-4), 327-351.

\reference
Ellsworth, W. L. (1995).
Characteristic earthquakes and long-term earthquake forecasts:
implications of central California seismicity,
in F. Y. Cheng and M. S. Sheu (Editors), {\sl Urban Disaster
Mitigation: The Role of Science and Technology}, Elsevier
Science Ltd., Oxford.

\reference
Ellsworth, W. L., M. V. Matthews, R. M. Nadeau, S. P.
Nishenko, P. A. Reasenberg, and R. W. Simpson (1999).
A physically based earthquake recurrence model for estimation
of long-term earthquake probabilities,
{\sl U.S. Geol.\ Surv.\ Open-File Rept.\ 99-522}.

\reference
Enescu, B., J. Mori, M. Miyazawa, and Y. Kano, 2009.
Omori-Utsu law $c$-values associated with recent
moderate earthquakes in Japan,
{\sl Bull.\ Seismol.\ Soc.\ Amer.}, {\bf 99}(2A), 884-891.

\reference
Feller, W., 1966.
{\sl An Introduction to Probability Theory and its
Applications}, {\bf 2}, J. Wiley, New York, 626~pp.

\reference
Felzer, K. R., T. W. Becker, R. E. Abercrombie, G. Ekstr\"om
and J. R. Rice, 2002.
Triggering of the 1999 $M_w$ 7.1 Hector Mine earthquake by
aftershocks of the 1992 $M_w$ 7.3 Landers earthquake,
{\sl J. Geophys.\ Res.}, {\bf 107}(B9), 2190,
doi:10.1029/2001JB000911.

\reference
Gutenberg, B. and Richter, C.F., 1944.
Frequency of earthquakes in California,
{\sl Bull.\ seism.\ Soc.\ Am.}, {\bf 34}, 185-188.

\reference
Hanks, T.C., 1992.
Small earthquakes, tectonic forces,
{\sl Science}, {\bf 256}, 1430-1432.

\reference
Hawkes, A.\ G., and L.\ Adamopoulos, 1973.
Cluster models for earthquakes - Regional comparisons,
{\sl Bull.\ Int.\ Statist.\ Inst.}, {\sl 45(3)}, 454-461.

\reference
Helmstetter, A., and D. Sornette, 2003.
Importance of direct and indirect triggered seismicity in the
ETAS model of seismicity,
{\sl Geophys.\ Res.\ Lett.}, {\bf 30}(11), 1576,
doi:10.1029/2003GL017670.

\reference
Helmstetter, A., Y. Y. Kagan, and D. D. Jackson, 2005.
Importance of small earthquakes for stress transfers
and earthquake triggering,
{\sl J. Geophys.\ Res.}, {\bf 110}(B5), B05S08,
doi:10.1029/2004JB003286, pp.~1-13.

\reference
Helmstetter, A., Y. Y. Kagan, and D. D. Jackson, 2006.
Comparison of short-term and time-independent earthquake
forecast models for southern California,
{\sl Bull.\ Seismol.\ Soc.\ Amer.}, {\bf 96}(1), 90-106.

\reference
Jackson, D. D., and Y. Y. Kagan, 2006.
The 2004 Parkfield earthquake, the 1985 prediction, and
characteristic earthquakes: lessons for the future,
{\sl Bull.\ Seismol.\ Soc.\ Amer.}, {\bf 96}(4B), S397-S409,
doi: 10.1785/0120050821.

\reference
Kagan, Y.~Y., 1982.
Stochastic model of earthquake fault geometry,
{\sl Geophys.\ J.\ Roy.\ astr.\ Soc.}, {\bf 71}(3), 659-691.

\reference
Kagan, Y.~Y., 1991.
Likelihood analysis of earthquake catalogues,
{\sl Geophys.\ J. Int.}, {\bf 106}(1), 135-148.

\reference
Kagan, Y.~Y., 1994.
Distribution of incremental static stress caused by
earthquakes,
{\sl Nonlinear Processes Geophys.}, {\bf 1}(2/3), 172-181.

\reference
Kagan, Y. Y., 2002a.
Aftershock zone scaling,
{\sl Bull.\ Seismol.\ Soc.\ Amer.}, {\bf 92}(2), 641-655.

\reference
Kagan, Y. Y., 2002b.
Seismic moment distribution revisited: II. Moment conservation
principle, {\sl Geophys.\ J. Int.}, {\bf 149}(3), 731-754.

\reference
Kagan, Y. Y., 2004.
Short-term properties of earthquake catalogs and models of
earthquake source,
{\sl Bull.\ Seismol.\ Soc.\ Amer.}, {\bf 94}(4), 1207-1228.

\reference
Kagan, Y.~Y., 2005.
Earthquake slip distribution: A statistical model,
{\sl J. Geophys.\ Res.}, {\bf 110}(B5), B05S11,
doi:10.1029/2004JB003280, pp.~1-15.

\reference
Kagan, Y. Y., 2007.
Earthquake spatial distribution: the correlation dimension,
{\sl Geophys.\ J. Int.}, {\bf 168}(3), 1175-1194.

\reference
Kagan, Y. Y., P. Bird, and D. D. Jackson, 2010.
Earthquake patterns in diverse tectonic zones of
the Globe,
{\sl Pure Appl.\ Geoph.} ({\sl The Frank Evison Volume}), {\bf
167}(6/7), 721-741, DOI: 10.1007/s00024-010-0075-3.

\reference
Kagan, Y. Y., and H. Houston, 2005.
Relation between mainshock rupture process and Omori's law for
aftershock moment release rate,
{\sl Geophys.\ J. Int.}, {\bf 163}(3), 1039-1048,
doi:10.1111/j.1365-246X.2005.02772.x

\reference
Kagan, Y.~Y., and D.~D.~Jackson, 1991.
Long-term earthquake clustering,
{\sl Geophys.\ J. Int.}, {\bf 104}(1), 117-133.

\reference
Kagan, Y.~Y. and D.~D.~Jackson, 1999.
Worldwide doublets of large shallow earthquakes,
{\sl Bull.\ Seismol.\ Soc.\ Amer.}, {\bf 89}(5), 1147-1155.

\reference
Kagan, Y. Y. and Jackson, D. D., 2010.
Global earthquake forecasts,
{\sl Geophys.\ J. Int.}, accepted.

\reference
Kagan, Y.~Y., and L. Knopoff, 1987.
Random stress and earthquake statistics: Time dependence,
{\sl Geophys.\ J.\ Roy.\ astr.\ Soc.}, {\bf 88}(3), 723-731,
DOI: 10.1111/j.1365-246X.1987.tb01653.x.

\reference
Ken-Tor R, Agnon A, Enzel Y, Stein M, Marco S, Negendank JFW,
2001.
High-resolution geological record of historic earthquakes in
the Dead Sea basin,
{\sl J. Geophys.\ Res.}, {\bf 106}(B2), 2221-2234.

\reference
Kreemer, C., Holt, W. E., and Haines, A. J., 2003.
An integrated global model of present-day plate motions and
plate boundary deformation,
{\sl Geophys.\ J. Int.}, {\bf 154}, 8-34,
doi:10.1046/j.1365-246X.2003.01917.x

\reference
Lavall\'ee, D., 2008.
On the random nature of earthquake sources and ground motions:
a unified theory,
{\sl Advances in Geophysics}, {\bf 50}, 427-461.

\reference
Marsan, D., 2005.
The role of small earthquakes in redistributing crustal
elastic stress,
{\sl Geophys. J. Int.}, {\bf 163}(1), 141-151,
doi:10.1111/j.1365-246X.2005.02700.x

\reference
Marsan, D., and Lenglin\'e, O., 2008.
Extending earthquakes' reach through cascading,
{\sl Science},  {\bf 319}, 1076-1079.

\reference
Matthews, M. V., W. L. Ellsworth, and P. A. Reasenberg, 2002.
A Brownian model for recurrent earthquakes,
{\sl Bull.\ Seismol.\ Soc.\ Amer.}, {\bf 92}, 2233-2250.

\reference
McCann, W.\ R., S.\ P.\ Nishenko, L.\ R.\ Sykes, and J.\
Krause, 1979.
Seismic gaps and plate tectonics: seismic potential for major
boundaries,
{\sl Pure Appl.\ Geophys.}, {\bf 117}, 1082-1147.

\reference
Nadeau, R. M., W. Foxall, and T. V. McEvilly (1995).
Clustering and periodic recurrence of microearthquakes on the
San Andreas fault at Parkfield, California,
{\sl Science}, {\bf 267}, 503-507.

\reference
Nadeau, R. M., and Johnson, L. R., 1998.
Seismological studies at Parkfield VI: Moment release rates
and estimates of source parameters for small repeating
earthquakes,
{\sl Bull.\ Seismol.\ Soc.\ Amer.}, {\bf 88}, 790-814.

\reference
Nishenko, S. P., 1991.
Circum-Pacific seismic potential -- 1989-1999,
{\sl Pure Appl.\ Geophys.}, {\bf 135}, 169-259.

\reference
Ogata, Y., 1998.
Space-time point-process models for earthquake occurrences,
{\sl Ann.\ Inst.\ Statist.\ Math.}, {\bf 50}(2), 379-402.

\reference
Ogata, Y., Jones, L.M., Toda S., 2003.
When and where the aftershock activity was depressed:
Contrasting decay patterns of the proximate large earthquakes
in southern California,
{\sl J. Geophys.\ Res.}, {\bf 108}(B6), 2318.

\reference
Ogata, Y., and J. C. Zhuang, 2006.
Space-time ETAS models and an improved extension,
{\sl Tectonophysics}, {\bf 413}(1-2), 13-23.

\reference
Ojala, I. O., I. G. Main, and B. T. Ngwenya (2004).
Strain rate and temperature dependence of Omori law scaling
constants of AE data: Implications for earthquake
foreshock-aftershock sequences,
{\sl Geophys. Res. Lett.}, {\bf 31}, L24617,
doi:10.1029/2004GL020781

\reference
Omori, F., 1894.
On the after-shocks of earthquakes,
{\sl J.\ College Sci., Imp.\ Univ.\ Tokyo}, {\bf 7}, 111-200
(with Plates IV-XIX).

\reference
Parsons, T., 2009.
Lasting earthquake legacy,
{\sl Nature,} {\bf 462}(7269), 42-43.

\reference
Powers, P. M., and Jordan, T. H., 2010.
Distribution of seismicity across strike-slip faults in
California,
{\sl J. Geophys.\ Res.}, {\bf 115}(B05), Article Number:
B05305.

\reference
Rong, Y.-F., D. D. Jackson and Y. Y. Kagan, 2003.
Seismic gaps and earthquakes,
{\sl J. Geophys.\ Res.}, {\bf 108}(B10), 2471, {\bf ESE-6},
pp.~1-14, doi:10.1029/2002JB002334.

\reference
Rockwell, T. K., Lindvall, S., Herzberg, M., Murbach, D.,
Dawson, T., and G. Berger, 2000.
Paleoseismology of the Johnson Valley, Kickapoo, and Homestead
Valley faults: clustering of earthquakes in the eastern
California shear zone,
{\sl Bull.\ Seismol.\ Soc.\ Amer.}, {\bf 90}, 1200-1236.

\reference
Seshadri, V., 1999.
{\sl Inverse Gaussian Distribution: Statistical
Theory and Applications},
{\sl Lecture Notes in Statistics}, {\bf 137},
New York, Springer, 347~pp.

\reference
Stein, S., and M. Liu, 2009.
Long aftershock sequences within continents and implications
for earthquake hazard assessment,
{\sl Nature,} {\bf 462}(7269), 87-89.

\reference
U.S.\ Geological Survey, 2010.
{\sl Preliminary determinations of epicenters (PDE)},
(last accessed November 2010),
http://neic.usgs.gov/neis/epic/epic.html and
http://neic.usgs.gov/neis/epic/code\_catalog.html.

\reference
Utsu, T., Y. Ogata, and R. S. Matsu'ura, 1995.
The centenary of the Omori formula for a decay law of
aftershock activity,
{\sl J. Phys.\ Earth}, {\bf 43}, 1-33.

\reference
Wells, D. L., and K. J. Coppersmith, 1994.
New empirical relationships among magnitude, rupture length,
rupture width, rupture area, and surface displacement,
{\sl Bull.\ Seismol.\ Soc.\ Amer.}, {\bf 84}, 974-1002.

\reference
Wolfram, S., 1999.
{\sl The Mathematica Book},
4th ed., Champaign, IL, Wolfram Media, Cambridge, New York,
Cambridge University Press, pp.~1470.

\reference
Zaiser, M., 2006.
Scale invariance in plastic flow of crystalline solids,
{\sl Advances Physics}, {\bf 55}(1-2), 185-245.

\reference
Zhuang, J.C., Chang, C.-P., Ogata, Y., and
Chen, Y.-I., 2005.
A study on the background and clustering seismicity
in the Taiwan region by using point process models
{\sl J. Geophys.\ Res.}, {\bf 110}, B05S18,
doi:10.1029/2004JB003157.

\reference
Zolotarev, V.\ M., 1986.
{\sl One-Dimensional Stable Distributions},
Amer.\ Math.\ Soc., Providence, R.I., pp.~284;
Russian original 1983.

\clearpage

\newpage

\renewcommand{\baselinestretch}{1.25}

\begin{table}
% 2010/10/25 FPSM2L_PAIRS2.FOR;278 96/96 25-OCT-2010 18:07:08.75
\caption{Pairs of shallow earthquakes $m \ge 7.5$}
%\begin{planotable}{rrrrrrrrrrrrrrrrrr}
\vspace{10pt}
%\vspace{1pt}
\label{Table2}
%\begin{tabular}{rrrrrrrrrrrrrrrrrr}
\begin{tabular}{rrrrrrrrrrrrr}
%               123456789012345678
\hline
%& & & & & & & & & & & & \\[-25pt]
& & & & & & & & & & & & \\[-15pt]
% 2 3 4 5 6 7 8 9 0 1 2
\multicolumn{1}{c}{}&
\multicolumn{4}{c}{First Event}&
\multicolumn{4}{c}{Second Event}&
\multicolumn{3}{c}{Difference}&
\multicolumn{1}{c}{}
\\[2pt]
\cline{2-5}
\cline{6-9}
\cline{10-12}\\[-1.6ex]
%\cline{12-13}
\multicolumn{1}{c}{No}&
\multicolumn{1}{c}{Date}&
\multicolumn{2}{c}{Coord.}&
\multicolumn{1}{c}{$m$}&
\multicolumn{1}{c}{Date}&
\multicolumn{2}{c}{Coord.}&
\multicolumn{1}{c}{$m$}&
\multicolumn{1}{c}{$R$}&
\multicolumn{1}{c}{$\Phi$}&
\multicolumn{1}{c}{$\Delta t$}&
\multicolumn{1}{c}{$\eta$}\\[.3ex]
\multicolumn{2}{c}{}&
\multicolumn{1}{c}{Lat.}&
\multicolumn{1}{c}{Long.}&
\multicolumn{2}{c}{}&
\multicolumn{1}{c}{Lat.}&
\multicolumn{1}{c}{Long.}&
\multicolumn{1}{c}{}&
\multicolumn{1}{c}{km}&
\multicolumn{1}{c}{$^\circ$}&
\multicolumn{1}{c}{day}&
\multicolumn{1}{c}{}
\\[2pt]
\hline
& & & & & & & & & & & & \\[-15pt]
1 & 1977/06/22 & -22.9 & -174.9 & 8.1 & 2009/03/19 & -23.1 & -174.2 & 7.7 & 75 & 55 & 11593.26 & 1.4 \\
2 & 1978/03/23 & 44.1 & 149.3 & 7.6 & 1978/03/24 & 44.2 & 149.0 & 7.6 & 25 & 7 & 1.69 & 2.3 \\
3 & 1980/07/08 & -12.9 & 166.2 & 7.5 & 1980/07/17 & -12.4 & 165.9 & 7.8 & 62 & 18 & 8.85 & 1.1 \\
4 & 1980/07/08 & -12.9 & 166.2 & 7.5 & 1997/04/21 & -13.2 & 166.2 & 7.8 & 33 & 42 & 6130.53 & 2.0 \\
5 & 1980/07/08 & -12.9 & 166.2 & 7.5 & 2009/10/07 & -12.6 & 166.3 & 7.7 & 37 & 13 & 10682.95 & 1.6 \\
6 & 1980/07/17 & -12.4 & 165.9 & 7.8 & 2009/10/07 & -12.6 & 166.3 & 7.7 & 41 & 14 & 10674.10 & 1.8 \\
7 & 1980/07/17 & -12.4 & 165.9 & 7.8 & 2009/10/07 & -11.9 & 166.0 & 7.9 & 65 & 12 & 10674.11 & 1.3 \\
8 & 1983/03/18 & -4.9 & 153.3 & 7.8 & 2000/11/16 & -4.6 & 152.8 & 8.1 & 83 & 72 & 6452.83 & 1.3 \\
9 & 1983/03/18 & -4.9 & 153.3 & 7.8 & 2000/11/16 & -5.0 & 153.2 & 7.9 & 47 & 91 & 6452.94 & 1.8 \\
10 & 1985/09/19 & 17.9 & -102.0 & 8.0 & 1985/09/21 & 17.6 & -101.4 & 7.6 & 71 & 14 & 1.51 & 1.3 \\
11 & 1987/03/05 & -24.4 & -70.9 & 7.6 & 1995/07/30 & -24.2 & -70.7 & 8.1 & 33 & 7 & 3068.83 & 2.8 \\
12 & 1990/04/18 & 1.3 & 123.3 & 7.7 & 1991/06/20 & 1.0 & 123.2 & 7.6 & 37 & 29 & 427.65 & 1.6 \\
13 & 1995/08/16 & -5.5 & 153.6 & 7.8 & 2000/11/16 & -5.0 & 153.2 & 7.9 & 76 & 74 & 1918.89 & 1.1 \\
14 & 1997/04/21 & -13.2 & 166.2 & 7.8 & 2009/10/07 & -12.6 & 166.3 & 7.7 & 70 & 30 & 4552.42 & 1.0 \\
15 & 2000/06/04 & -4.7 & 101.9 & 7.9 & 2007/09/12 & -3.8 & 101.0 & 8.6 & 150 & 85 & 2655.78 & 1.3 \\
16 & 2000/11/16 & -4.6 & 152.8 & 8.1 & 2000/11/16 & -5.0 & 153.2 & 7.9 & 67 & 89 & 0.12 & 1.7 \\
17 & 2000/11/16 & -4.6 & 152.8 & 8.1 & 2000/11/17 & -5.3 & 152.3 & 7.8 & 93 & 88 & 1.67 & 1.2 \\
18 & 2001/06/23 & -17.3 & -72.7 & 8.5 & 2001/07/07 & -17.5 & -72.4 & 7.7 & 34 & 8 & 13.54 & 4.7 \\
19 & 2005/03/28 & 1.7 & 97.1 & 8.7 & 2010/04/06 & 2.0 & 96.7 & 7.8 & 58 & 7 & 1835.25 & 4.0 \\
20 & 2006/11/15 & 46.7 & 154.3 & 8.4 & 2007/01/13 & 46.2 & 154.8 & 8.2 & 70 & 82 & 58.71 & 2.7 \\
21 & 2007/09/12 & -3.8 & 101.0 & 8.6 & 2007/09/12 & -2.5 & 100.1 & 7.9 & 176 & 11 & 0.53 & 1.2 \\
22 & 2007/09/12 & -3.8 & 101.0 & 8.6 & 2010/10/25 & -3.7 &  99.3 & 7.9 & 189 & 8 & 1139.15 & 1.1 \\
% 1            2       3        4     5            6       7        8     9    0    1          2
\hline
\end{tabular}

%\end{planotable}

\bigskip
$R$ -- centroid distance,
$\Phi$ -- 3-D rotation angle between focal mechanisms,
$\Delta t$ -- time interval between events,
$\eta$ -- degree of zone overlap, the ratio of earthquake
focal zone sizes to twice their distance, see Equations (2,3)
in Kagan and Jackson (1999).
The total earthquake number with magnitude $m \ge 7.50$
for the period 1976/1/1--2010/10/25 is 121.
The maximum epicentroid distance is 250.00~km.

\hfil\break
\vspace{5pt}
\end{table}

\newpage

\begin{table}
% 2010/09/21  P53,pp.69-70 BIRD_ZONES.FOR;HARPDE5D2.FOR; SCAL1A1.M;SCAL1C.M
\caption{Aftershock zone log length vs mainshock moment magnitude $m$}
%\begin{planotable}{rrrrrrrrrrrrrrrrrr}
\vspace{10pt}
%\vspace{1pt}
\label{Table1}
\begin{tabular}{rccccrccr}
\hline
& & & & & & & &  \\[-15pt]
\multicolumn{1}{c}{\#}&
\multicolumn{1}{c}{Tectonic zone}&
\multicolumn{1}{c}{Focal mech.}&
\multicolumn{1}{c}{$a_0$}&
\multicolumn{1}{c}{$a_1$}&
\multicolumn{1}{c}{$a_2$}&
\multicolumn{1}{c}{$\sigma$}&
\multicolumn{1}{c}{$\epsilon_{{\rm max}}$}&
\multicolumn{1}{c}{$n$}
\\[2pt]
\hline
& & & & & & & &  \\[-15pt]
1 & All & All & 2.48 & 0.492 & -- & 0.134 & 0.468 & 160 \\
2 & All & All & 2.48 & 0.493 & 0.0013 & 0.134 & 0.468 & 160 \\
& & & & & & & &  \\[-12pt]
3 & All & Thrust & 2.48 & 0.501 & -- & 0.132 & 0.457 & 115 \\
4 & All & Thrust & 2.48 & 0.499 & 0.0022 & 0.132 & 0.458 & 115 \\
& & & & & & & &  \\[-12pt]
5 & All & Normal & 2.47 & 0.532 & -- & 0.076 & 0.132 & 15 \\
6 & All & Normal & 2.46 & 0.427 & $-0.0884$ & 0.075 & 0.137 & 15 \\
& & & & & & & &  \\[-12pt]
7 & All & Str.-Slip & 2.47 & 0.437 & -- & 0.153 & 0.276 & 30 \\
8 & All & Str.-Slip & 2.49 & 0.490 & 0.0038 & 0.153 & 0.278 & 30 \\
& & & & & & & &  \\[-12pt]
& & & & & & & &  \\[-12pt]
9 & Trench & All & 2.47 & 0.499 & -- & 0.131 & 0.454 & 129 \\
10 & Trench & All & 2.47 & 0.482 & $-0.0177$ & 0.131 & 0.449 & 129 \\
& & & & & & & &  \\[-12pt]
11 & Trench & Thrust & 2.48 & 0.500 & -- & 0.135 & 0.460 & 104 \\
12 & Trench & Thrust & 2.48 & 0.488 & $-0.0142$ & 0.135 & 0.455 & 104 \\
& & & & & & & &  \\[-12pt]
13 & Trench & Normal & 2.47 & 0.543 & -- & 0.067 & 0.120 & 12 \\
14 & Trench & Normal & 2.46 & 0.440 & $-0.0915$ & 0.065 & 0.114 & 12 \\
& & & & & & & &  \\[-12pt]
15 & Trench & Str.-Slip & 2.38 & 0.409 & -- & 0.146 & 0.302 & 13 \\
16 & Trench & Str.-Slip & 2.29 & 0.034 & $-0.2790$ & 0.142 & 0.268 & 13 \\
& & & & & & & &  \\[-12pt]
& & & & & & & &  \\[-12pt]
17 & Active Cont. & All & 2.52 & 0.504 & -- & 0.138 & 0.256 & 27 \\
18 & Active Cont. & All & 2.74 & 1.180 & 0.4250 & 0.135 & 0.220 & 27 \\
& & & & & & & &  \\[-12pt]
19 & Active Cont. & Thrust & 2.44 & 0.480 & -- & 0.112 & 0.176 & 10 \\
20 & Active Cont. & Thrust & 3.00 & 2.210 & 1.0800 & 0.064 & 0.130 & 10 \\
& & & & & & & &  \\[-12pt]
21 & Active Cont. & Str.-Slip & 2.50 & 0.419 & -- & 0.145 & 0.314 & 14 \\
22 & Active Cont. & Str.-Slip & 2.65 & 0.889 & 0.2990 & 0.143 & 0.295 & 14 \\
& & & & & & & &  \\[-12pt]
\hline
\end{tabular}

%\end{planotable}

\bigskip
$a_0$, $a_1$, $a_2$ are regression coefficients in
Eq.~\ref{Eq18}; $\sigma$ is the standard uncertainty;
$\epsilon_{{\rm max}}$ is the maximum error;
$n$ is the number of aftershock sequences.
The catalogs time interval is 1977/1/1--2010/09/21.

\hfil\break
\vspace{5pt}
\end{table}

\newpage

\clearpage

\renewcommand{\baselinestretch}{1.75}

\parindent=0mm

% \newpage

\newpage

\end{document}